\documentclass[american,aps,prx,reprint,floatfix,superscriptaddress,longbibliography]{revtex4-1}

\usepackage{graphicx}
\usepackage{dcolumn}
\usepackage{bm}
\usepackage{amsmath}
\usepackage{amssymb}
\usepackage{physics}
\usepackage{bbm}
\usepackage[dvipsnames]{xcolor}
\usepackage[capitalise]{cleveref}
\usepackage{floatrow}
\usepackage{cprotect}
\usepackage{tabularx}
\usepackage[labelfont=bf,labelformat=simple]{subfig}

\definecolor{color1}{HTML}{648FFF}
\definecolor{color2}{HTML}{DC267F}
\definecolor{color3}{HTML}{FFB000}

\begin{document}
\floatsetup[figure]{style=plain,subcapbesideposition=top}
\title{Empirical learning of dynamical decoupling on quantum processors}
\author{Christopher Tong}
\email{lctong@mit.edu}
 \affiliation{\vspace{-1pt}Department of Physics, Massachusetts Institute of Technology, Cambridge, MA, USA}
 \affiliation{\vspace{-1pt}IBM Quantum, IBM Thomas J. Watson Research Center, Yorktown Heights, NY, USA}
 
\author{Helena Zhang}
\email{helena.zhang@ibm.com}
\affiliation{\vspace{-1pt}IBM Quantum, IBM Thomas J. Watson Research Center, Yorktown Heights, NY, USA}

\author{Bibek Pokharel}
\email{bibek.pokharel@ibm.com}
\affiliation{\vspace{-1pt}IBM Quantum, IBM Thomas J. Watson Research Center, Yorktown Heights, NY, USA}

\date{\today}

\begin{abstract}
Dynamical decoupling (DD) is a low-overhead method for quantum error suppression. Despite extensive work in DD design, finding pulse sequences that optimally decouple computational qubits on noisy quantum hardware is not well understood. In this work, we describe how learning algorithms can empirically tailor DD strategies for any quantum circuit and device. We use a genetic algorithm-inspired search to optimize DD (GADD) strategies for IBM's superconducting-qubit based quantum processors. In all observed experimental settings, we find that empirically learned DD strategies significantly improve error suppression relative to canonical sequences, with relative improvement increasing with problem size and circuit sophistication. We leverage this to study mirror randomized benchmarking on 100 qubits, GHZ state preparation on 50 qubits, and the Bernstein-Vazirani algorithm on 27 qubits. We further demonstrate that our empirical learning method finds strategies, in time constant with increasing circuit width and depth, that provide stable performance over long periods of time without retraining and generalize to larger circuits when trained on small sub-circuit structures.
\end{abstract}

\maketitle
\section{Introduction}
\label{sec:intro}

Dynamical decoupling (DD)~\cite{viola1999dynamical,Viola:98,Vitali:99,Zan99} was one of the earliest methods to be proposed for quantum error suppression~\cite{carr1954effects,meiboom1958modified}, which is vital for useful quantum computation before fault tolerance~\cite{Sho95, Ste96, Sho96, Pre18} and can continue to improve quantum computation in the fault-tolerant regime~\cite{Ngh11}. Due to the low overhead of applying control pulses during qubit idle periods to suppress errors, there have been extensive theoretical studies and experimental demonstrations of DD across superconducting circuits \cite{pokharel2018demonstration,Ezz22,Jurcevic_2021, Goo23, Kim23, Kim23_2}, trapped ions \cite{Bie09}, and solid state electron spins \cite{Duj09,de2010universal,Wan12, farfurnik2016improving}. 

Numerous novel DD sequences have been proposed in the past two decades. These sequences have focused on specific error suppression goals, such as the cancellation of a given term in the system-bath interaction Hamiltonian, increasing the order in the pulse spacing $\tau$ to which errors are suppressed, or reducing the effect of systematic errors in pulse implementation on a single qubit. We refer curious readers to Ref.~\cite{Ezz22} for a comprehensive survey of various DD sequences. The scale of quantum circuits used in today's experiments has grown beyond that of existing theoretical approaches. In particular, single-qubit sequences such as Carr-Purcell-Meiboom-Gill (CPMG), universally robust DD (URDD), and Eulerian DD (EDD) \cite{Gen17,Vio03} are often applied to multi-qubit experiments~\cite{singkanipa2024demonstration,Pok22,Pok23} despite the fact that theoretical guarantees for these sequences in canceling single-qubit errors do not extend to multi-qubit quantum crosstalk~\cite{Tri22}, a central source of error in large circuits. 

Staggered single-qubit DD sequences have been studied recently for crosstalk suppression~\cite{Tri22, Zho23, Shi24} on larger circuits because while multi-qubit DD sequences exist, their application on quantum processors rarely succeeds~\cite{Qui24} as multi-qubit quantum gates are substantially more error-prone than single-qubit ones~\cite{krantz2019quantum}. But DD sequence staggering also introduces its own complications. Finding sequences that cancel overlapping crosstalk terms in a quantum circuit becomes progressively non-trivial as qubit number and circuit complexity increases, reflective of the challenge presented by describing the noise characteristics of a device in the context of quantum control restrictions due to circuit structure~\cite{krantz2019quantum,zhang2022predicting}. 

To address these challenges in DD optimization, we leverage classical optimization techniques, which have broadly enhanced quantum computational results to enable cutting-edge experiments~\cite{kandala2019error, Kim23_2, van2023probabilistic, temme2017error,nation2021scalable,van2022model,pokharel2024scalable,yang2022efficient}. Classical methods such as deep learning~\cite{Zlo20} and variational methods~\cite{Das21, Rav22} have been applied to study DD, but have only provided marginal improvements. We extend the use of classical optimization for DD by providing a framework to empirically learn DD strategies tailored to any given quantum device and task at hand. Our specific choice of the learning algorithm builds on the work in Ref.~\cite{Qui13}, where the authors use a genetic algorithm to improve single-qubit quantum memory by considering a classical simulation of open quantum system dynamics. We generalize this approach by including empirical feedback from multi-qubit circuits executed on bona fide quantum devices. We give a systematic description of how to train DD sequences for a given quantum circuit, even when it is large and classically intractable to simulate. 

We go beyond detailing our empirical DD learning strategy, a genetic algorithm-inspired search to optimize DD (GADD), by demonstrating its capabilities across three experimental settings. We apply GADD for quantum error suppression in an oracular algorithm, a state preparation circuit, and a randomized benchmarking circuit ensemble. In each case, we show that empirically trained sequences comfortably outperform canonical sequences. We also report the extension of mirrored randomized benchmarking (MRB) to 100 qubits, which was possible through the application of GADD but not with any canonically studied DD strategies. Our experimental results provide compelling demonstrations of the scalability, robustness, and generalizability of empirical DD learning.

We start by reviewing DD in Sec.~\ref{sec:background_dd} with a focus on the search space of possible strategies to motivate our empirical scheme. In Sec.~\ref{sec:methods}, we discuss the specific implementation of our learning scheme. In Sec.~\ref{sec:results}, we describe the results achieved by our method in an experimental setting. Finally, in Sec.~\ref{sec:conc}, we discuss the implications of our results for DD design and the broader study of noisy quantum computation on large quantum circuits.

\section{Background on Dynamical Decoupling}
\label{sec:background_dd}
We begin with a simplified framework of a noisy system that is typical of the context where DD sequences are theoretically analyzed. Consider an idle period of qubit evolution where the system dynamics are governed by a time independent system-bath interaction Hamiltonian $H_{SB}$ and bath-specific Hamiltonian $H_B$. For time $\tau$, the system evolution is given by the unitary operator
\begin{equation}
    f_\tau = \exp[- i \tau (H_{SB} + H_{B})].
\end{equation}
Let us consider the decoupling group $G \subseteq \mathrm{SU}(2)$ such that \(g_{j} \in G\) represent physical actions on the system Hilbert space \(\mathcal{H}_S\). $G$ acts on the set of $f_\tau$ by conjugation:
\begin{align}
\begin{split}
    g_j^\dagger f_\tau g_j &= \exp[- i \tau g_j^\dagger(H_{SB} + H_{B})g_j] \\ 
    &= \exp[- i \tau (H_{SB}' + H_{B}')],
\end{split}
\end{align}
where $H_{SB}' = g_j^\dagger H_{SB} g_j$.  As an example, consider the  general single-qubit system-bath coupling Hamiltonian, expressed as a sum of tensor products between Pauli operators and bath operators:
\begin{equation}
    H_{SB} = \sum\limits_{\alpha=x,y,z} \sigma^{\alpha} \otimes  B^{\alpha},
\end{equation}
which implies that
\begin{equation}
    H_{SB}' = \sum\limits_{\alpha=x,y,z} \left(\sigma^\alpha + [g_j^\dagger, \sigma^\alpha]g_j\right)\otimes B^{\alpha}.
    \label{eqn:hsbprime}
\end{equation}
We consider the concatenation of $L-1$ such intervals of length $\tau$ conjugated by possibly distinct $g_j \in G$. The corresponding unitary evolution operator over time $(L-1)\tau$ is then
\begin{align}
\begin{split}
    U(T) &= \prod_{j=1}^{L-1} g_{j}^{\dagger} f_{\tau} g_{j} \\
    &= \exp[- i\tau \sum_{j=1}^{L-1} (H_{SB, j}' + H_B) + O(\tau^2)]
\end{split}
\end{align}
by the BCH formula, where $H_{SB,j}'$ is the effective Hamiltonian as derived in \cref{eqn:hsbprime} due to conjugation by the $j^{\text{th}}$ element of the decoupling group present in the sequence. Such a time interval where the system evolves under $H_{SB}$ conjugated by elements of $G$ can be experimentally achieved by the length $L$ pulse sequence given by $p_1=g_1^\dag$, $p_{L} = g_{L-1}$ and $p_j = g_{j-1}g_{j}^{\dag}$ for $j = 1, \dots, L-1$ at the corresponding times in the pulse sequence. As such, $G$ can be chosen such that the pulses $p_j$ are easily implemented from existing operations present on the quantum system. Thus, performing DD requires nothing more than applying a pre-programmed sequence of pulses during idle periods in any quantum computation. For example, taking $g_1 = X$ from $G= \{I, X\}$ results in the $L=2$ pulse sequence $p_1 = p_2 = X$ and cancels all terms $H_{SB}$ not coupled to $\sigma^x$ to first order in $\tau$, while taking $g_1 = X$, $g_2 = Z$, $g_3 = Y$ from $G = \{I, X, Y, Z\}$ results in the XY4 pulse sequence which cancels all system-bath interactions up to first order in the pulse spacing $\tau$~\cite{Ezz22}. 

Although the selective cancellation of system-bath interactions with DD can be much more sophisticated, as exemplified by dynamically generated decoherence-free subspaces \cite{Zan99, Wul02, Qui24}, XY4-like sequences that provide maximal suppression of system-bath interactions to first order have been of particular theoretical interest~\cite{Lid14}. There is a theoretical equivalence class of DD sequences achieved by selecting a permutation of the non-identity elements of $G$ to be the $\{g_j\}$, of which there are $(L-1)!$. For instance, $XYXY$ belongs to the same theoretical equivalence class as $XZXZ$.~\footnote{For a decoupling group of size $\abs{G}$ and pulse sequences of length $L$, there are $\binom{\abs{G}+L-2}{\abs{G}-1}$ equivalence classes of DD sequences with distinct $U(T)$ operators when neglecting contributions to the effective system-bath Hamiltonian of $O(\tau^2)$ and higher. This result arises from counting the number of ways that the $L-1$ intervals where the system evolves under the effective Hamiltonian $g_j^\dag H_{SB} g_j$ can be partitioned among the $\abs{G}$ possibilities for $g_j$.} However, the equivalence of such sequences can be broken due to experimental gate implementations. For example, $X$ and $Y$ gates are often implemented as finite width control pulses on quantum hardware, while $Z$ is implemented as a virtual state rotation, leading to distinct experimental behavior of theoretically equivalent DD sequences~\cite{pokharel2018demonstration,Ezz22}. Additionally, it is important to look beyond the theoretically optimal equivalence class of single qubit DD sequences as the decoupling group includes the identity $I$, which does not behave as a DD pulse in the traditional sense but serves to control the gap lengths between pulses. Finally, in a many-qubit circuit, $C > 1$ independent DD sequences can be simultaneously applied to different circuit idle periods. 

As a result, the overall space of DD strategies considered  for optimization has size $\abs{G}^{(L-1)C}$, where each $g_j$ can independently take on any element of $G$. In this paper, we show that it is possible to systematically search this large space and empirically identify well-performing DD strategies. Moreover, when performing our search, we often find sets of DD sequences that are empirically equivalent in their decoupling power with respect to the experimental constraints and imperfections, even when it is hard to discern if there are structural similarity between these sequences. 

\begin{figure*}
    \centering
    \includegraphics[width = 18cm]{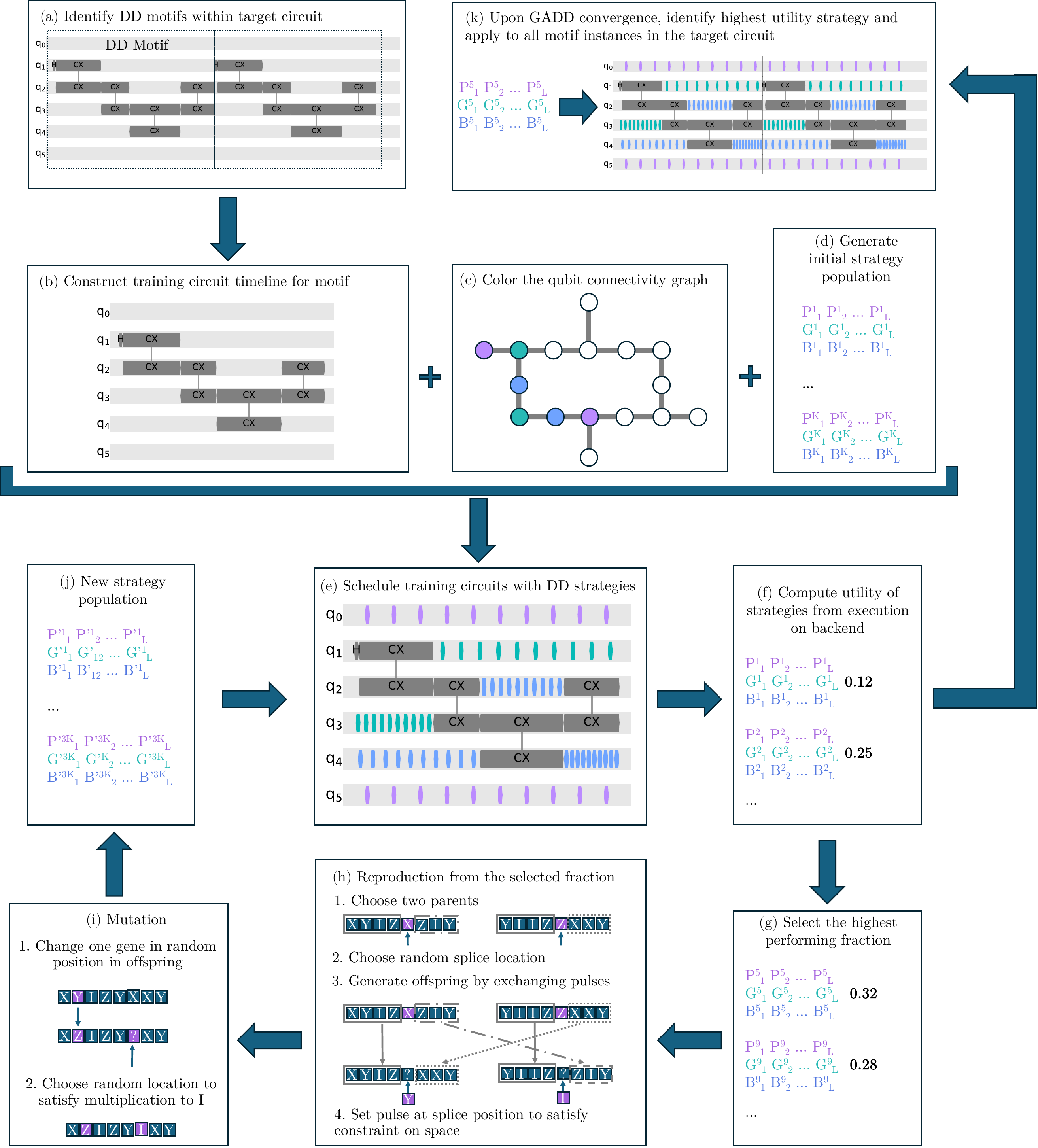}
    \caption{Flowchart for searching effective DD strategies using the genetic algorithm. After (a) identifying the motif and (b) laying out the corresponding circuit timeline, select qubits of the device are (c) 3-colored into $c=\{\text{P (Purple), G (Green), B (Blue)}\}$. Each individual in the initial population of size $K$ is (d) associated with a PGB DD sequence that is (e) scheduled on the circuit. After utility functions are (f) computed via empirical execution on the backend, they are used to (g) select individuals for (h) reproduction. Utility functions are then recursively computed on the (j) resulting population of size $3K$ after applying (i) mutations to select the new GA population. At any point, iteration can be terminated and the highest utility DD strategy (k) selected for experimental application.}
    \label{fig:gadd_flowchart}
\end{figure*}

\section{Methods \label{sec:methods}}
The GADD framework for empirically tailoring DD for a given quantum computer and target quantum circuit with classical learning algorithms is depicted in Fig.~\ref{fig:gadd_flowchart}. In Sec.~\ref{ss:training}, we describe the construction of the utility function and computationally efficient training circuits as inputs for empirical learning, corresponding to Fig.~\ref{fig:gadd_flowchart}(a)-(b) and Fig.~\ref{fig:gadd_flowchart}(k). The learning algorithm itself, corresponding to Fig.~\ref{fig:gadd_flowchart}(c)-(j), is described in Sec.~\ref{ss:genetic}. 

\subsection{Training for a target circuit \label{ss:training}}
Given the task of finding optimal DD strategies for a given \textit{target circuit}, our method relies on the careful selection of a \textit{utility function} and \textit{training circuit} such that upon execution of the training circuit with any given DD strategy, the utility function applied on the resulting population distribution is an effective measure of how well the DD strategy suppresses errors on the target circuit. Importantly, the training and target circuits need not be identical. 

In the simplest case, the target circuit deterministically marks a correct state over all computational basis states upon measurement. In such cases, we let the training circuit to be identical to the target circuit and set the utility function $f_U$ to be the success probability $p_s$:
\begin{equation}
    f_U = p_s = \frac{\text{Number of correct state measurements}}{\text{Number of total shots}}
    \label{eqn:sp_utility}
\end{equation}
We demonstrate this in Sec.~\ref{sec:bv} by applying GADD to the Bernstein-Vazirani problem as a first benchmark of our empirical DD search. 

In a more general setting, if the ideal outcome probability distribution $\vb{p}(k)$ over computational basis states of a quantum circuit is known, then the utility metric can be generalized from the success probability to the 1-norm between $\vb{p}(k)$ and observed $\hat{\vb{p}}(k)$ is used as the utility function:

\begin{equation}
    f_U = 1 - \frac{1}{2} \sum_{k=0}^{2^N-1} \abs{\vb{p}(k) - \hat{\vb{p}}(k)}
    \label{eqn:1norm_utility}
\end{equation}
In Sec.~\ref{sec:ghz}, we demonstrate empirical DD optimization for GHZ state generation using this utility function. 

However, $\vb{p}(k)$ is not necessarily known for general application-scale quantum circuits, which are often hard to simulate classically. To address this challenge, we carefully construct training circuits with easily computable outcome distributions such that high-utility DD strategies on training circuits yield effective error suppression for the target circuit. We achieve this by identifying sub-circuit building blocks, which we call \textit{DD motifs}, on which we independently optimize DD. Motifs can be as simple as an idle gap on a single qubit, which the bulk of the literature in DD optimization has focused on~\cite{Ezz22}, or as complex as the entire target circuit, as discussed in the previous cases. Generally, the selection of training circuits that lie between these two extrema relies on the assumption that optimal DD strategies depend on circuit structures but not specific gate-level parameters. This is well established through prior work, such as DD studies on the motifs of neighboring idle qubits on the device graph~\cite{Zho23}, qubits adjacent to an entangling operation~\cite{Tri22}, and generic 4-qubit Clifford sub-circuits~\cite{Das21}. Our outlined method provides an efficient pathway for DD optimization on general motifs, which is necessary for efficient empirical DD optimization for generic quantum circuits with arbitrary circuit structures.

We build training circuits to simplify target circuits and arrive at easily computable utility functions while
preserving identified DD motifs. For instance, DD motifs that are repeated multiple times in the target circuit can be used only once in the training circuit. Most importantly, we highlight two straightforward strategies for simplifying target circuit motifs in a general setting---circuit mirroring and Cliffordization. In circuit mirroring as proposed in Ref.~\cite{Pro22_2}, for a given motif, the training circuit is constructed by appending the inverse of each gate present in the motif in reverse order, maintaining the timings of all idle gaps, with symmetry breaking-gates added in between the motif and its inverse. Such a circuit is efficiently simulable as the motif composed with its inverse behaves as the identity under noiseless time evolution. In Cliffordization, we take the motif of interest and replace all non-Clifford gates with their Clifford counterparts, a generalization of the 4-qubit DD optimization proposed in Ref.~\cite{Das21}. As the Clifford group acts as the stabilizer of the group of all unitary transformations on $N$ qubits~\cite{Aar04}, the ideal outcome distribution of a Cliffordized circuit is efficiently simulable. In both cases, Eqn.~\ref{eqn:sp_utility} acts as a suitable utility function of the training circuit output to learn DD strategies for the target circuit. To demonstrate the effectiveness of training circuits that substitute individual gates in a DD motif while preserving its structure, we demonstrate experiments on two problems of interest: mirror randomized benchmarking in Sec.~\ref{sec:mrb} and Grover's algorithm in Appendix~\ref{ss:clifford}.

\subsection{Genetic algorithms \label{ss:genetic}}
Given a training circuit and a utility function, we must also define a search space of potential DD strategies and a way to navigate this space, i.e. the learning algorithm. The DD strategy space we consider is the aforementioned space of $C$ different length $L$ pulse sequences drawn from a given decoupling group $G$ on an $N$-qubit training circuit. Throughout this work, we refer to each of the $C$ distinct pulse sequences as corresponding to a distinct \textit{color}. In order to search this large discrete space, we employ a genetic algorithm, an approach inspired by natural selection to optimize a given utility function~\cite{reeves2002genetic,Mit99}. In the prototypical genetic algorithm, an individual in a population can be identified by features analogous to genes on a chromosome. By identifying two individuals in the population and a scheme to exchange features among the two, additional individuals of the population can be generated in a process analogous to reproduction. Mutations can also be induced on individuals to increase the population diversity.

The problem of determining DD strategies for a quantum circuit admits an unambiguous representation in the context of a genetic algorithm~\cite{Qui13}; in particular, the single qubit pulses $\{p_j\}$ making up each length $L$ sequence for each of the $C$ colors are treated as the specific genes that identify unique elements of the population of DD strategies. The structure of the decoupling group $G$ enforces that all $p_j \in G$; in this paper, we consider pulses from the group
\begin{equation}
    G = \{I_p, I_m, X_p, X_m, Y_p, Y_m, Z_p, Z_m\}
\end{equation}
where $I_p, X_p, Y_p, Z_p$ are the pulses associated with the typical Pauli matrices and $I_m, X_m, Y_m, Z_m$ are their respective equivalents with rotation in the opposite direction on the Bloch sphere~\footnote{$I_m$ and $I_p$ both act as the identity element, although $I_m$ is included for purposes of group structure}. Most well-known canonical DD sequences, such as XY4, EDD, and CPMG and its $X_pX_m$ variant, are within this search space. The sequence length $L$ is chosen such that idle periods in the target circuit are sufficiently long for the padding of $L$ pulses. Finally, we partition the $N$ circuit qubits into $C$ sets by coloring a graph where nodes represent circuit qubits and edges represent entangling gates among qubits. Our coloring scheme was chosen to minimize the effect of crosstalk, but DD coloring structure can be more general. For instance, the colors need not be restricted to coloring graph nodes and instead each idle gap for each qubit could be assigned a different color. We found experimentally that $C$ can be much smaller than $N$ for achieving high degrees of error suppression as long as each color corresponds to a qubit set where no pair of qubits experiencing an entangling gate receive the same DD sequence (Fig.~\ref{fig:gadd_flowchart}(c)). For all $1 \leq c \leq C$, idle gaps experienced by all qubits in the $c^{\text{th}}$ partition receive the $c^{\text{th}}$ DD sequence in any given strategy. For the experimental results herein, we use $C \leq 3$ as the IBM quantum processor's heavy hexagon architecture is planar and cannot contain any subgraph isomorphic to $K_4$ \cite{Cha20}. An analogous process can be performed on other limited-connectivity architectures. The chosen coloring is performed to explore a sequence space including generalized staggered DD sequences for suppression of qubit-qubit crosstalk~\footnote{To impose that such allocations of identity operations indeed lead to distinct staggerings, the DD sequence associated with the first color in the strategy has the pulse sequence placed uniformly between the operations defining the start and end of the idle period, while the second receives an asymmetric sequence with a pulse at the earliest time one can be applied in the idle period and the third receives an asymmetric sequence with a pulse at the latest time one can be applied in the idle period.}.

We construct an initial population of $K$ DD strategies (Fig.~\ref{fig:gadd_flowchart}(d)), which are then inserted onto corresponding idle gaps in the given quantum circuit (Fig.~\ref{fig:gadd_flowchart}(e)). The resulting circuit is executed on the quantum computer for a fixed number of shots. For each strategy, we compute a utility function of the observed distribution via projective measurement of the final state given by the circuit (Fig.~\ref{fig:gadd_flowchart}(f)). The $K$ highest utility strategies are selected to create the next strategy generation (Fig.~\ref{fig:gadd_flowchart}(g)), where $K$ pairs of individuals are selected with replacement for reproduction to produce $2K$ offspring (Fig.~\ref{fig:gadd_flowchart}(h)), which are then subject to mutation (Fig.~\ref{fig:gadd_flowchart}(i)). The circuits of the combination of parents and offspring of size $3K$ are then executed. Iteration continues (Fig.~\ref{fig:gadd_flowchart}(j)) either until a satisfactorily high utility DD strategy is found or a set number of iterations is reached (Fig.~\ref{fig:gadd_flowchart}(k)).

We selected the genetic algorithm as the base for our approach due to the inherent parallelization in numerous independently evolving DD strategies associated with different colors which can be evaluated simultaneously.  Our choice of discrete optimization over a continuous optimization scheme was deliberate. DD sequences can be described by many continuous parameters, such as the rotation axis and angle, duration, and timing of each pulse and such parameters can be fed into continuous parameter optimization algorithms \cite{Pol2012}, such as gradient-based methods or alternative heuristic methods (e.g. Nelder-Mead), in a similar fashion to pulse-based variational strategies \cite{Egg23, Rav22}. In particular, stochastic gradient descent is used in Ref.~\cite{Rah24} to optimize a single DD sequence on a Bell state preparation circuit containing up to eight intermediate qubits. However, a continuous approach introduces a large number of parameters, and devising a principled approach to reduce these parameters for a tractable number of quantum-to-classical calls to convergence on large-scale quantum circuits is difficult. Our discrete optimization method provides reliable and efficient convergence to high-utility DD strategies on large circuits without incurring the detrimental effects of exponential runtime blowup.

\section{Results and Discussion \label{sec:results}}
Here we highlight the use of GADD on a variety of problems. In each instance, an empirical DD search yields a positive impact on computational results. These experiments were chosen to exemplify three different choices for the utility function as well as GADD training settings: success probability of the target circuit for the Bernstein-Vazirani algorithm  (Sec.~\ref{sec:bv}), which is an oracular algorithm; 1-norm of the observed probability distribution for the GHZ state preparation circuit (Sec.~\ref{sec:ghz}); success probability of smaller sub-circuits from a representative random sample for mirror randomized benchmarking (MRB), a large-scale benchmarking protocol for programmable quantum computers (Sec.~\ref{sec:mrb}).

\subsection{Bernstein-Vazirani algorithm \label{sec:bv}}

\begin{figure*}[t]    
      \includegraphics[height=5.5cm]{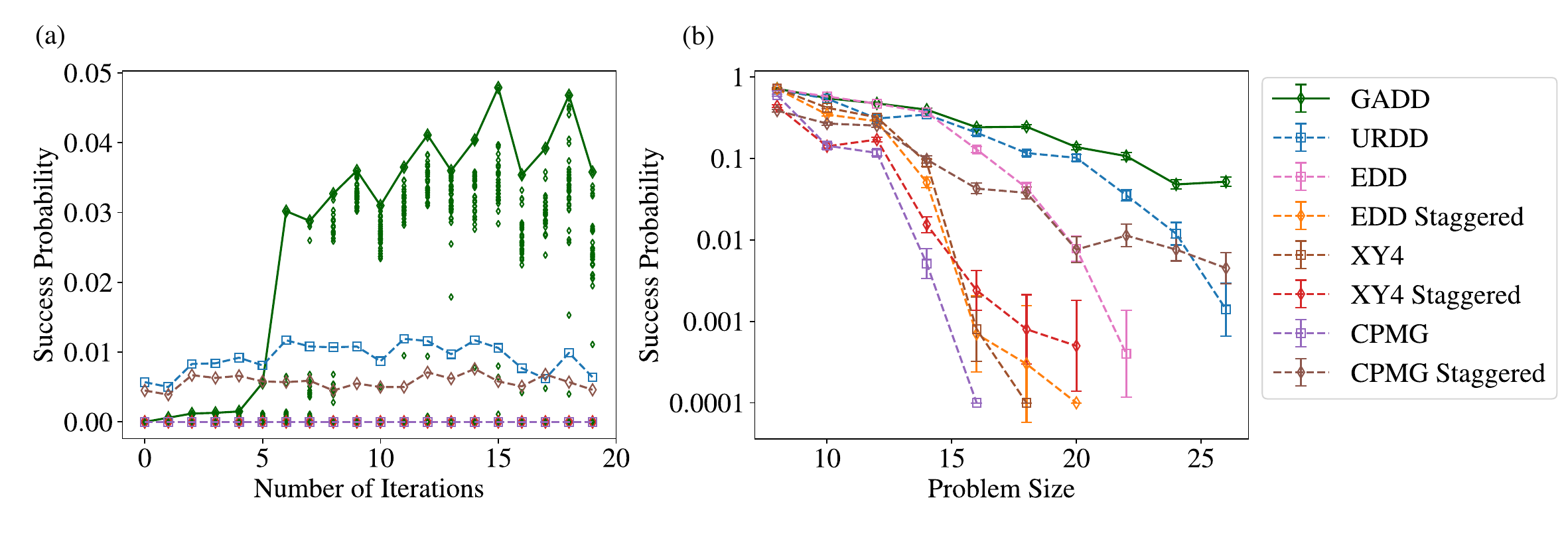}
    \caption{Success probability for the Bernstein-Vazirani circuit on $\texttt{ibm\_peekskill}$ (a) on problem size 24 for one training session and (b) as a function of problem size for BV-8 to BV-26. GADD is compared against a suite of established DD methods. Staggered sequences are represented by open diamonds and non-staggered sequences by open squares. Fluctuations of $\sim1\%$ in the success probability are observed due to systematic device performance fluctuations between GA iterations. The maximum success probability of finding the $1^n$ bitstring is evaluated across different values of $n$ and GADD represents a significant improvement over all other tested DD methods.}
    \label{fig:bv_main}
\end{figure*}
The Bernstein-Vazirani (BV) algorithm was one of the first oracular quantum algorithms to demonstrate a theoretical quantum speedup \cite{Ber93}. In its conventional form, BV provides a linear quantum speedup over classical computation, and recently, an algorithmic quantum speedup was demonstrated in single-shot BV \cite{Pok23} via the use of URDD for error suppression. The existence of positive results with existing DD methods, as well as the linear scaling of the BV quantum circuit with problem size and a deterministic target string, makes BV a standard choice for benchmarking protocols on programmable quantum devices ~\cite{lubinski2023application,fallek2016transport,wright2019benchmarking}. Here, we use the BV quantum circuit to compare the performance of GADD against a suite of well-established DD sequences commonly used in experimental settings today, including the symmetric and staggered versions of CPMG, XY4, and EDD, as well as the UR$_{16}$ universally robust sequence. In these staggered versions, qubits of different colors experience staggered pulse timings; URDD is not staggered as the URDD protocol selects pulse numbers based on idle duration lengths and thus has its own internal timing regulation \cite{qiskit_research}.

BV-$n$ corresponds to the quantum task implementing an oracle encoding hidden bitstring $b \in \{0,1\}^n$, which outputs $f_b(x) = b \cdot x \pmod{2}$ when queried with the input bitstring $x$. The goal is to identify $b$ with the minimal number of oracle queries. The classical success probability after one iteration is $p_{\text{s,classical}} = 2/2^n$ while the ideal quantum success probability is $p_{\text{s,quantum}}=1$ and requires $N=n+1$ qubits to implement. While there are $2^n$ different oracles in the BV-$n$ problem class, we choose to consider only $b=1^n$ to be representative of the problem for each $n$ as it has the largest circuit depth, following the lead of Refs.~\cite{Pok23,lubinski2023application,Mun23}. These circuits are run on the 27-qubit \verb|ibm_peekskill| backend with the IBM Falcon architecture. We consider the problem sizes $n = 8, 10, 12, \dots, 26$ up to the largest possible size on a 27-qubit device and repeat each circuit for 10,000 shots. The GADD experiment is performed with sequences with length $L = 8$ for $20$ iterations and initial population with size $K = 16$ with Eqn.~\ref{eqn:sp_utility} as the utility function with $1^n$ as the correct state. 

First, we focus our attention on the convergence of the GADD algorithm in the $n = 24$ case, which is characteristic of the GADD behavior on the BV problem in the larger $n$ regime (Fig.~\ref{fig:bv_main}a). We consider the highest success probability over strategies in the population at each iteration versus the iteration number to demonstrate the convergence of the GADD population to one with maximum success probability $0.04$ within $10$ iterations. This represents more than a $4$-fold increase in the success probability achieved with the URDD sequence and staggered CPMG, while all other sequences give success probability $p_s=0$. 

The general case in the large $n$ regime yields similar results upon comparing GADD with benchmark sequences. We find that our empirical method strikingly outperforms all theoretically derived sequences in the test suite (Fig.~\ref{fig:bv_main}b). For comparison, in the smaller problem size regime of $n\lessapprox 16$, we find that both GADD and URDD have high success probabilities, reproducing previous work that empirically demonstrated good error suppression with URDD~\cite{Ezz22,Pok23,singkanipa2024demonstration}. The relative improvement of empirically learned GADD strategies over URDD increases with problem size, and the success probability obtained via GADD achieves a much slower decay of $p_s$ with respect to problem size $n$ when compared to all other DD strategies in our test suite. 

We further note some particularly interesting results from the DD test suite. The staggered DD sequences did not generally improve in performance with respect to their non-staggered counterparts. Though staggering of the sequence timing is successful in the regime of aligned idle periods between qubits, as explained in Ref.~\cite{Zho23}, for idle periods that are coincidentally misaligned by half of the time difference between adjacent DD pulses, the equivalent staggered structure for crosstalk cancellation is achieved by imposing equal pulse timings as fractions of the gap time during both idle gaps. When the number of gap-pairs between adjacent device qubits grows in an unstructured application circuit, we reach the regime where the distribution of optimal spacings for each gap-pair spans the entire staggering spectrum. In this limit, we do not \textit{a priori} expect the naive application of a specific pulse staggering to significantly outperform all others due to crosstalk cancellation across the entire circuit. Another interesting result was that robust DD sequences such as EDD did not necessarily exhibit higher computational fidelities than non-robust sequences such as CPMG and XY4. This is not necessarily unexpected as these sequences are only robust under certain single-qubit pulse imperfections, like flip-angle errors, which may not be the most prominent error source for the chosen device. On the other hand, URDD was constructed to be robust to a broader variety of errors, has the inherent pulse timing selection functionality, and performed quite well with respect to the other canonical methods in the suite. 

Although it may be possible that there can be further improvement in the DD performance with continuous pulse spacing optimization or further hyperparameter tuning in GADD or other methods such as URDD, such improvement necessitates a computational tradeoff. Additionally, improvement in computational fidelity due to DD is ultimately limited by the fact that all DD sequences are only effective against non-Markovian errors~\cite{lidar2013quantum}. Most pertinently, the ability of GADD to learn DD strategies that match and exceed the performance of sequences in the test suite, each of which resulted from insightful theoretical analyses, highlights that DD design for physical systems can be outsourced to empirical learning algorithms.

\subsection{GHZ state preparation and persistence analysis\label{sec:ghz}}

\begin{figure*}[t]    
      \includegraphics[height = 5cm]{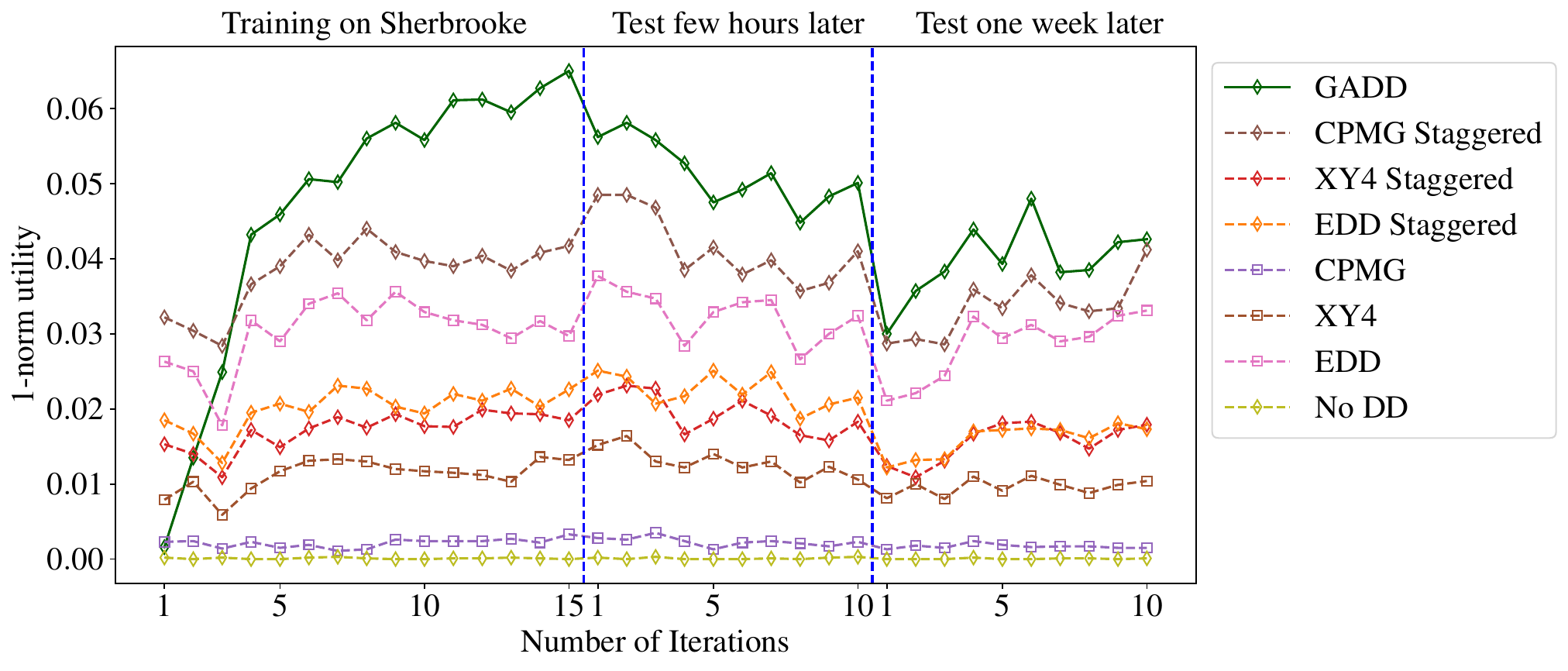}
    \cprotect\caption{Persistence of learned sequences over time. Empirical learning on the 50-qubit GHZ state preparation circuit \verb|ibm_sherbrooke| finds DD strategies that outperform all canonical sequences in our test suite within 4 iterations. After a few hours, the noise environment on the quantum device does not experience significant changes and the relative ordering of DD performance is preserved. In particular, application of the previously learned GADD strategies leads error suppression beyond that provided by canonical sequences without incurring any additional overhead. After one week, fluctuations in the device noise profile lead to slight deterioration in GADD strategy performance, but the GADD sequences are still competitive and can form the starting point for the next set of training iterations.}
    \label{fig:time_pers}
\end{figure*}

\begin{figure*}[t]    
      \includegraphics[height = 5cm]{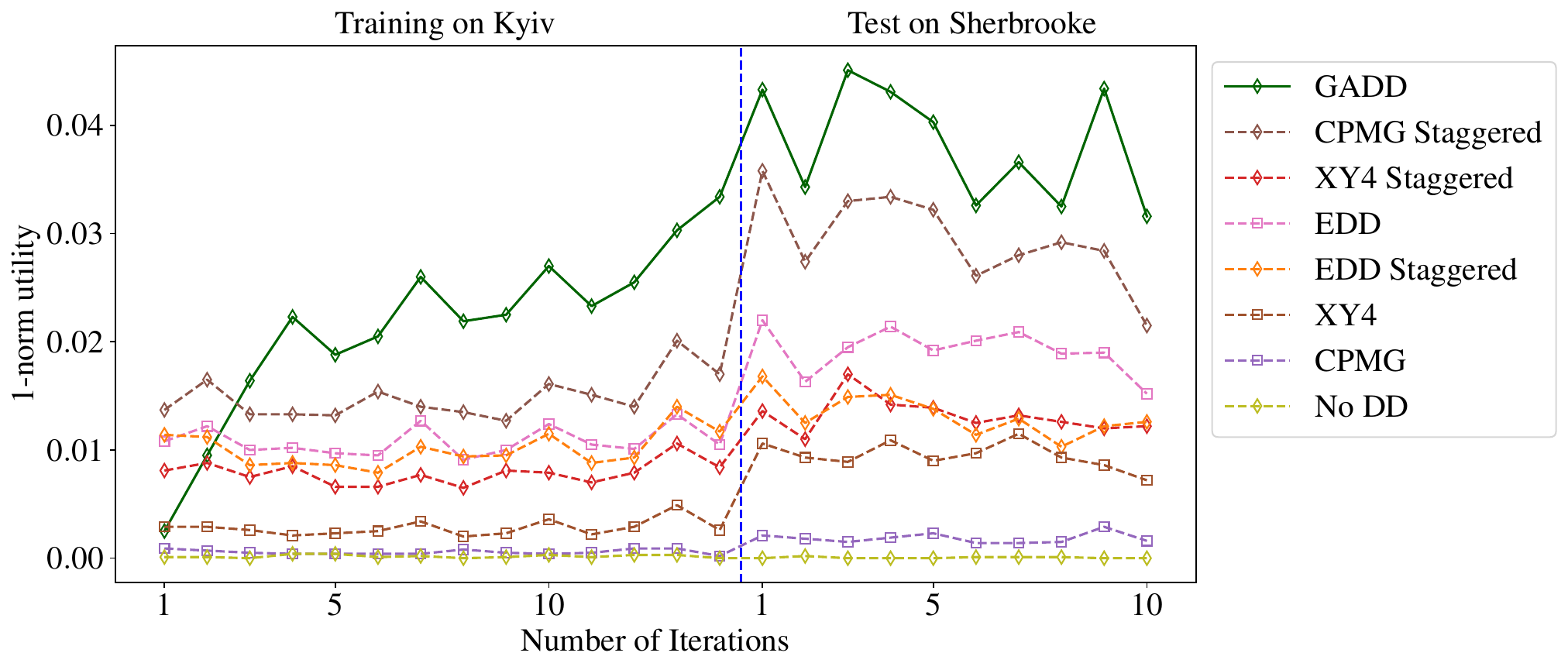}
    \cprotect\caption{Transferability of empirically learned sequences between devices. When empirically learning on the 50-qubit GHZ state preparation circuit on the \verb|ibm_kyiv| device, we find GADD strategies that outperform all canonical sequences in our test suite within 4 iterations. Upon transferring the GADD strategy population on the same circuit to the \verb|ibm_sherbrooke| device, we see very similar relative performance, demonstrating the generalizability of GADD-learned strategies across superconducting devices of the same architecture.}
    \label{fig:device_pers}
\end{figure*}

The Greenberger-Horne-Zeilinger (GHZ) state on $n$ qubits, \(\frac{1}{\sqrt{2^n}} \left( \ket{0}^{\otimes n} + \ket{1}^{\otimes n} \right)\), is an entangled state of great relevance in quantum computation and communication~\cite{Gre89, Hil99}. Preparing the GHZ state is often the first step for numerous applications such as encoding for quantum error-correcting codes~\cite{Omk22, Pan24}, quantum cryptography~\cite{Che08}, and Hamiltonian simulation~\cite{Son19,Che23}. Consequently, error suppression strategies to improve their preparation are of interest. We prepare a 50-qubit GHZ state to analyze the performance of empirically learned DD strategies for the preparation of these large, highly entangled states. As effectively preparing large and high-fidelity GHZ states en masse is a standard routine of important quantum applications, we also test the persistence and robustness of GADD strategies over time compared with canonical sequences in order to determine the extent to which empirically learned DD sequences can be re-applied for many instances of GHZ state preparation without incurring the overhead of additional learning.

For GADD training on GHZ states, we take the 1-norm utility function (Eqn.~\ref{eqn:1norm_utility}), where the theoretically expected distribution is given by
$$ \vb{p}(k) = \begin{cases}
0.5 & \text{if } k \in \{0^n, 1^n\} \\
0 & \text{otherwise}
\end{cases} $$
and the training circuit is taken to be identical to the 50-qubit GHZ state generation target circuit. The GADD experiment is first run for 15 iterations in order to empirically learn DD strategies for GHZ state generation. To test the persistence of these learned sequences over time, the final surviving population of the genetic algorithm is then tested for 10 iterations, without any further optimization, a few hours later, and the same test is repeated with the new final population a week later. All experiments were performed on the same set of qubits on \verb|ibm_sherbrooke|. Results are depicted in Fig.~\ref{fig:time_pers}. In all cases, the empirically learned DD sequences perform significantly better than all tested canonical sequences and the relative DD sequence rankings remain consistent over time. We note that the difference between the empirically learned sequence and the best-performing canonical sequence, CPMG staggered, decreases over time. This is consistent with the expectation that DD sequences tailored to the device at a specific time will have decreased performance when device specifications change, shifting the landscape of single-qubit errors, multi-qubit crosstalk, and any other non-Markovian errors present. However, the remaining performance improvement between empirically learned and canonical DD sequences provide evidence that GADD learns to suppress many aspects of the noise environment that generalize across time, providing effectively zero application overhead to apply GADD  subsequent to the initial learning experiment. 

To further test the robustness of the learned sequences, we perform the learning experiment on one device and test the result on a different programmable device with the same underlying architecture using the same training circuit and utility function. We again perform 15 GADD iterations for empirical DD learning and apply the final population of learned sequences on \texttt{ibm\_kyiv} to \texttt{ibm\_sherbrooke}, on which we perform test the learned sequences for 10 repetitions of the experiment. The results are shown in Fig.~\ref{fig:device_pers}. Once again, the empirically discovered DD sequences outperform all the evaluated standard sequences and the relative rankings of these DD sequences remain consistent. As such, we demonstrate the transferability of empirically learned DD sequences across devices. This further supports the efficiency of DD learning, as it is not necessary to perform subsequent learning experiments for effective DD performance across devices of the same architecture.

\subsection{Mirror randomized benchmarking \label{sec:mrb}}

\begin{figure}[h!t]
    \includegraphics[width = \textwidth]{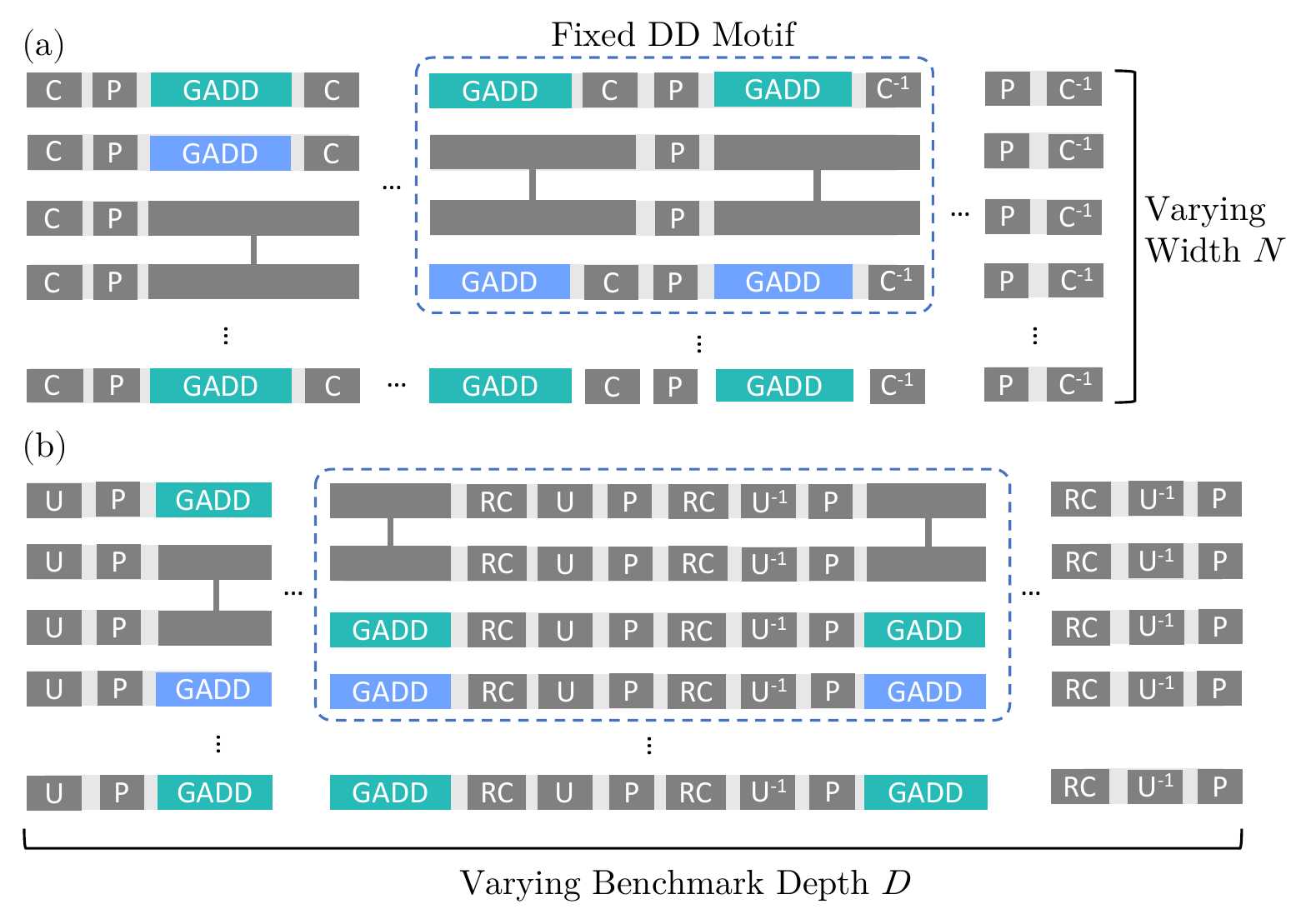}

    \caption{Training and target circuits for (a) Clifford and (b) universal mirror randomized benchmarking \cite{Pro22, Hin23}. $D/2$ layers of one-qubit gates (\textsf{C} for Cliffords and \textsf{U} for SU(2)) and two-qubit gate layers are mirrored across both circuits with corresponding inverses (\textsf{C$^{-1}$} and \textsf{U$^{-1}$}) for a total of $D$ benchmark layers, and each has temporal gaps from the difference in one- and two-qubit gate times that enable the insertion of short GADD sequences. Uniformly random layers of Paulis (\textsf{P}) dress the benchmark layers, and the universal MRB single-qubit layers are further dressed with gates from randomized compiling (\textsf{RC}). While the overall circuits vary in the number of qubits $N$ and $D$, training motifs are based on the structure of smaller fixed circuits (dotted boxes) extracted from the larger circuits and randomized to be distinct from the target circuits.}
    \label{fig:mrb_procedure}
\end{figure}

Mirror randomized benchmarking (MRB) is a scalable variant of the standard randomized benchmarking (RB) \cite{Mag2011, Mag2012} protocol that characterizes the noise of quantum processors by measuring the error rates of randomized layers sampled from a defined gate set \cite{Pro22, Hin23}. In Clifford MRB \cite{Pro22}, $D$ benchmark layers of single-qubit Clifford gates and two-qubit gates are inverted and reflected across the central time axis of the circuit. Layers of random Pauli gates dress each of the benchmark layers and a single-qubit Clifford layer and its inverse are applied at the start and end of the protocol respectively (\cref{fig:mrb_procedure}). In the universal version of the protocol that uses gates sampled from a universal gate set such as $\mathrm{SU}(2)$, composite layers are constructed by randomly sampling single-qubit unitary gates from the designated gate set with a specified density of two-qubit entangling gates, and the second half of the circuit is the inverse of the first half randomly compiled to prevent systematic error cancellation \cite{Hin23}. In either protocol, mirror circuits are run on a processor and then an effective polarization $S$, a weighed alternative to success probability, is calculated for each $N$-qubit circuit as derived in Ref.~\cite{Pro22}:
\begin{equation}
S = \frac{4^N}{4^N - 1}\left[\sum_{k=0}^N\left(-\frac12\right)^k h_k\right] - \frac{1}{4^N -1},
\end{equation}
where $h_k$ is the probability that the circuit output bitstring is Hamming distance $k$ from the target bitstring.

The error per layer (EPL) is then calculated by fitting an exponential decay curve, $Ap^D$, to the relation of $S$ versus the benchmark depth $D$:

\begin{equation}
\text{EPL} = \frac{2^N-1}{2^N}\left(1-p\right),
\end{equation}

which corresponds to the average gate infidelity of the Pauli-dressed $N$-qubit benchmark layer.

\begin{figure*}[ht!]
    \includegraphics[width=\textwidth]{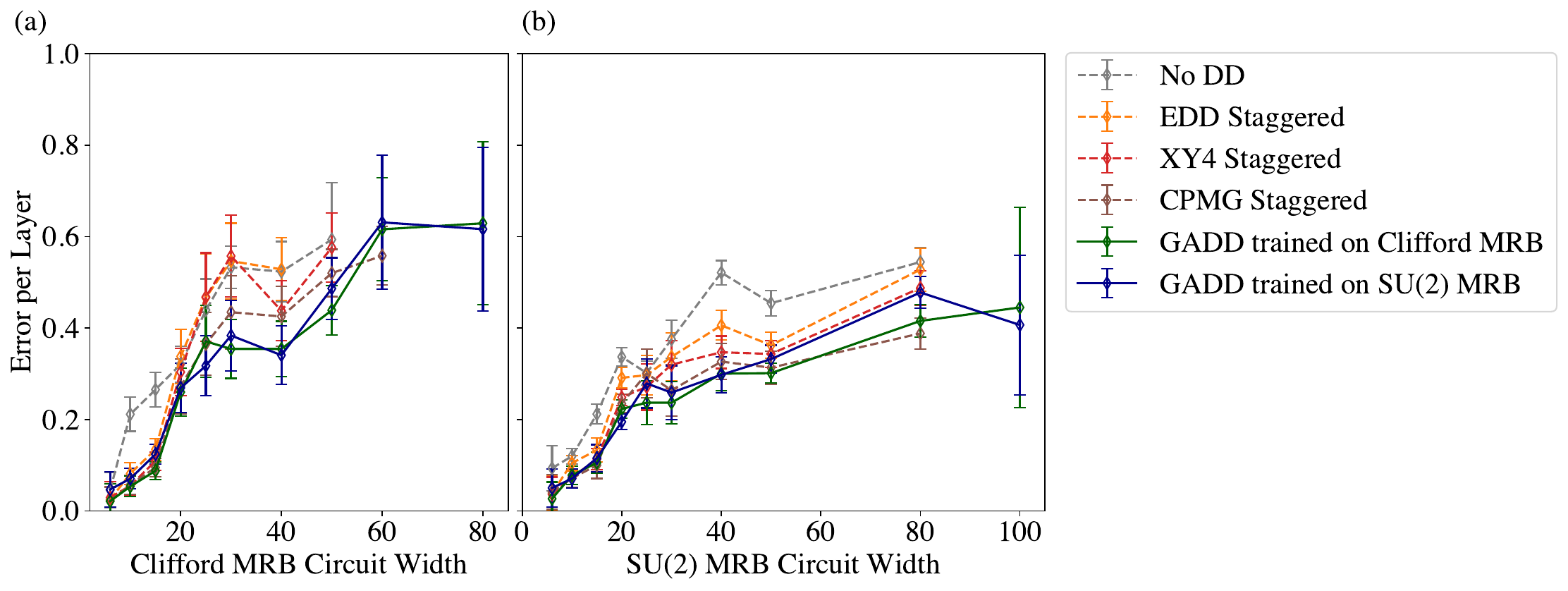}
    \caption{Mirror randomized benchmarking results on \texttt{ibm\_kyiv}. After training on a sample of small Clifford and $\mathrm{SU}(2)$ MRB circuit at a fixed number of circuit layers, the optimal GADD strategy is applied to target (a) Clifford MRB and (b) $\mathrm{SU}(2)$ MRB circuits and compared with canonical DD sequences as well as the bare circuits without DD by fitting measured polarizations at varying circuit widths and depths to obtain the EPL at each width. At larger numbers of qubits, not using DD or using suboptimal DD sequences results in polarization that cannot be fit to obtain an EPL; the data is shown for as long as an EPL can be obtained. Both GADD strategies outperform canonical DD sequences and are the only usable benchmarks at 80 qubits for the Clifford circuit and 100 qubits for the $\mathrm{SU}(2)$ circuit, such that MRB with GADD can be used to benchmark up to 100-qubit linear chains on this device.}
    \label{fig:mrb iterations}
    \label{fig:mrb results}
\end{figure*}

While MRB is more scalable than standard RB, which is infeasible beyond five qubits \cite{Pro22}, it still has difficulty scaling to the regime of $>50$ qubits on current superconducting hardware, where other benchmarks such as layer fidelity \cite{mckay2023} must be used instead. The difficulty is mainly due to crosstalk as the number of possible nontrivial crosstalk interactions contributing to decoherence increases with qubit number while the individual gates that MRB is intended to benchmark typically have similar error rates across an entire device, leading to roughly constant EPL per qubit in the crosstalk-free regime \cite{Hin23, Ami23}. As a result, there exists a threshold number of qubits for current hardware above which an effective polarization cannot be extracted even for the smallest circuit depths due to noise. We are therefore interested in examining whether it is possible to increase the scalability of MRB on real devices by reducing crosstalk using dynamical decoupling on randomized MRB circuits in the 50+ qubits regime. This would expand MRB's utility for benchmarking the infidelity of $N$-qubit layers on and across quantum processors.

MRB is a challenging and relevant test case for the efficacy of GADD for multiple reasons. First, since MRB exists in both Clifford and universal versions with similar, though not identical, circuit structures, it is a good candidate to test the validity of training on a motif that differs from the target circuit in both size and structure. If we can train on either Clifford or universal MRB circuits and find that the best GADD strategy performs well on the other problem as well, then it suggests that a technique like Cliffordization, where the training circuit is also not identical in timing to the target circuit, is likely to work on similar cases. Second, the nature of attempting to extend the benchmark to larger numbers of qubits necessitates using training circuits over smaller numbers of qubits, since the full-size target circuit will likely not have enough signal-to-noise to train on. Furthermore, the condensed idle periods between layers in a typical MRB circuit are sufficiently short such that longer robust sequences cannot be padded; the previous effort to improve the signal-to-noise used simple CPMG sequences \cite{Ami23}. Lastly, the randomized nature of MRB circuits implies the training and target circuits must be fully random and not share repeating motifs even for a constant number of qubits. Therefore, to succeed, GADD must find sequences that lower error rates on small motif circuits with relatively small idle gaps for DD insertion and also lower error rates on larger circuits with slightly different gates and timings due to the randomized nature of MRB circuits.

Taking all of the above points into account, we conduct two separate trainings, one on a single smaller motif of the Clifford MRB circuit and the other on the analogous motif of the SU(2) universal circuit, as shown in \cref{fig:mrb_procedure}, then apply the learned strategy to larger circuits of both Clifford and SU(2) MRB. We perform our GADD algorithm on strategies with $L = 4$ sequences across the 127-qubit \verb|ibm_kyiv| backend, which is a superconducting qubit device with the IBM Eagle architecture. We use subsets of linear qubit arrays, which only require two different colors alternating on adjacent qubits. To train the GADD algorithm, we calculated the utility function by computing the average success probability over a fixed set of $5$ random MRB circuits on 10 qubits containing $6$ circuit layers. We note that in addition to being a smaller circuit motif compared to the target circuit, this is significantly fewer circuits than what would be needed for a typical MRB experiment, which requires sampling sets of random circuits at different numbers of layers. The echoed cross-resonance gate \cite{malekakhlagh2020first}, equivalent to CNOT up to single-qubit pre-rotations, was used as the two-qubit gate of choice on the layers sampled with the edge grab sampler \cite{Pro22_2} with a two-qubit gate density of $0.25$. The convergence to the asymptotic maximal utility happened quickly within $10$ iterations, similar to previous experimental demonstrations. The DD strategy yielding the highest success probability after training was then padded across a suite of random MRB circuits at varying benchmark depths to fit the effective polarization to an exponential and extract the EPL. The target circuits were executed without DD and with canonical XY4, EDD, and CPMG sequences, with timing staggered between the two qubit colorings, along with the application of both GADD sequences for comparison. At each circuit width of interest, we use the exact same sub-array for all experimental runs.

The results are shown in \cref{fig:mrb results}. While Clifford MRB circuits beyond 50 qubit linear strings without dynamical decoupling already fail to produce any useful infidelity signal and some canonical sequences are able to reduce EPL to a certain extent, the best GADD strategies learned via training on Clifford and SU(2) circuits are both demonstrably lower in EPL and capable of extending the range to 80 qubits~(\cref{fig:mrb results}(a)). The increased range also corresponds to reduced average error per layer as DD effectively reduces crosstalk contributions. This improvement extends to $\mathrm{SU}(2)$ MRB, where we observe that both circuits without DD and with canonical DD register lower EPL figures, but the GADD sequences are the only ones that extend to 100 qubits~(\cref{fig:mrb results}(b)). In Ref.~\cite{Hin23}, an experimental demonstration on a 4-qubit linear array of a superconducting processor showed comparable error per layer rates from both Clifford and $\mathrm{SU}(2)$ MRB, which is similar to our results at lower numbers of qubits. It is only past $\sim$20 qubits where the lower error rates from $\mathrm{SU}(2)$ MRB becomes evident. We hypothesize that the $\mathrm{SU}(2)$ MRB infidelities are broadly lower than the corresponding Clifford MRB numbers because the randomized gates themselves have decoupling properties~\cite{Lea06}, and randomly sampling gates from $\mathrm{SU}(2)$ provides more effective decoupling than sampling gates from a subset such as the Clifford group.

Our MRB results demonstrate the effectiveness of GADD on randomized circuits even when training on motif circuits far smaller than the desired target circuit and its potential for extending the scalability of such protocols, including, but not limited to, $\mathrm{SU}(2)$ universal MRB. There is no significant difference between the performance between the strategies learned from Clifford and SU(2) circuits, which suggests that as long as the circuit structure of the motif resembles that of the final circuit, it is possible for the GADD training procedure to find a DD strategy that achieves results with lower errors and greater scalability beyond using canonical DD sequences. It shows that Cliffordization is a promising technique for using GADD on target circuits with universal gatesets different from purely Clifford training circuits. The strategy of using GADD to extend the range of effectiveness of the MRB benchmark can also be applied more broadly to other such characterization tools.

\section{Conclusion \label{sec:conc}}

It is well known that error suppression on simple circuit structures, such as aligned periods of idle evolution on a block of qubits, can be optimized. However, deriving optimal sequences for general quantum circuits is highly non-trivial. As a result, while a plethora of increasingly sophisticated DD sequences exist in the literature, most experimental-scale quantum computing experiments only use simple DD sequences, such as XY4 or CPMG. Here, we provide a substantial extension of previous ideas in DD optimization~\cite{Qui24,Tri22,Das21} by creating a framework to systematically and efficiently search the space of DD strategies. With the large size and relatively unknown structure of the DD strategy space, matching the performance of well-studied robust sequences by empirical optimization is in itself a worthy goal. Here, we show that our learned sequences not only match but consistently outperform canonical DD strategies. Furthermore, we demonstrate that the performance of these better-than-canonical DD sequences persist over the passage of time and the transfer between devices of similar architecture, decreasing the overhead associated with DD learning. Finally, we demonstrate the generality of our method by taking small Cliffordized motifs from large-scale MRB circuits for DD training to extend the useful range for universal MRB on a programmable quantum computer to 100 qubits. 

Studying the population of optimized strategies determined by empirical DD learning can also be used to learn noise profiles and pulse imperfections of quantum devices. By considering patterns in the decoupling behavior and inherent robustness of learned sequences, we can determine the relative importance of cancelling specific errors on a quantum device. For example, our method allows for users to determine if strategies that outperform uniformly, multi-axis decoupling sequences such as XY4 indeed exist on their application of choice. In each learning context that we have considered, we find that the learned sequences almost always perform multi-axis decoupling with variations in pulse timing. Comprehensive quantitative analysis of the error suppression power of our identified sequences is left to future work. We hope that our work will inspire further work in using optimization techniques, perhaps beyond genetic algorithms, for navigating the large space of DD sequences to optimize error suppression. We are optimistic that such future work can be applied to discover new DD sequences of theoretical interest and acquire new insights about error suppression strategies and the physical errors affecting quantum devices.

\section*{Acknowledgements}

We acknowledge Luke Govia for helpful discussions and Francisco Rilloraza and Albert Zhu for developing the software to run MRB experiments. CT acknowledges support from the Quantum Undergraduate Research at IBM and Princeton (QURIP) internship program. HZ acknowledges support from the Army Research Office under QCISS (W911NF-21-1-0002). The views and conclusions contained in this document are those of the authors and should not be interpreted as representing the official policies, either expressed or implied, of the Army Research Office or the U.S. Government. The U.S. Government is authorized to reproduce and distribute reprints for Government purposes notwithstanding any copyright notation herein.

\bibliography{main}

\begin{thebibliography}{83}%
\makeatletter
\providecommand \@ifxundefined [1]{%
 \@ifx{#1\undefined}
}%
\providecommand \@ifnum [1]{%
 \ifnum #1\expandafter \@firstoftwo
 \else \expandafter \@secondoftwo
 \fi
}%
\providecommand \@ifx [1]{%
 \ifx #1\expandafter \@firstoftwo
 \else \expandafter \@secondoftwo
 \fi
}%
\providecommand \natexlab [1]{#1}%
\providecommand \enquote  [1]{``#1''}%
\providecommand \bibnamefont  [1]{#1}%
\providecommand \bibfnamefont [1]{#1}%
\providecommand \citenamefont [1]{#1}%
\providecommand \href@noop [0]{\@secondoftwo}%
\providecommand \href [0]{\begingroup \@sanitize@url \@href}%
\providecommand \@href[1]{\@@startlink{#1}\@@href}%
\providecommand \@@href[1]{\endgroup#1\@@endlink}%
\providecommand \@sanitize@url [0]{\catcode `\\12\catcode `\$12\catcode
  `\&12\catcode `\#12\catcode `\^12\catcode `\_12\catcode `\%12\relax}%
\providecommand \@@startlink[1]{}%
\providecommand \@@endlink[0]{}%
\providecommand \url  [0]{\begingroup\@sanitize@url \@url }%
\providecommand \@url [1]{\endgroup\@href {#1}{\urlprefix }}%
\providecommand \urlprefix  [0]{URL }%
\providecommand \Eprint [0]{\href }%
\providecommand \doibase [0]{http://dx.doi.org/}%
\providecommand \selectlanguage [0]{\@gobble}%
\providecommand \bibinfo  [0]{\@secondoftwo}%
\providecommand \bibfield  [0]{\@secondoftwo}%
\providecommand \translation [1]{[#1]}%
\providecommand \BibitemOpen [0]{}%
\providecommand \bibitemStop [0]{}%
\providecommand \bibitemNoStop [0]{.\EOS\space}%
\providecommand \EOS [0]{\spacefactor3000\relax}%
\providecommand \BibitemShut  [1]{\csname bibitem#1\endcsname}%
\let\auto@bib@innerbib\@empty
\bibitem [{\citenamefont {Viola}\ \emph {et~al.}(1999)\citenamefont {Viola},
  \citenamefont {Knill},\ and\ \citenamefont {Lloyd}}]{viola1999dynamical}%
  \BibitemOpen
  \bibfield  {author} {\bibinfo {author} {\bibfnamefont {L.}~\bibnamefont
  {Viola}}, \bibinfo {author} {\bibfnamefont {E.}~\bibnamefont {Knill}}, \ and\
  \bibinfo {author} {\bibfnamefont {S.}~\bibnamefont {Lloyd}},\ }\bibfield
  {title} {\enquote {\bibinfo {title} {Dynamical decoupling of open quantum
  systems},}\ }\href@noop {} {\bibfield  {journal} {\bibinfo  {journal}
  {Physical Review Letters}\ }\textbf {\bibinfo {volume} {82}},\ \bibinfo
  {pages} {2417} (\bibinfo {year} {1999})}\BibitemShut {NoStop}%
\bibitem [{\citenamefont {Viola}\ and\ \citenamefont {Lloyd}(1998)}]{Viola:98}%
  \BibitemOpen
  \bibfield  {author} {\bibinfo {author} {\bibfnamefont {Lorenza}\ \bibnamefont
  {Viola}}\ and\ \bibinfo {author} {\bibfnamefont {Seth}\ \bibnamefont
  {Lloyd}},\ }\bibfield  {title} {\enquote {\bibinfo {title} {Dynamical
  suppression of decoherence in two-state quantum systems},}\ }\href@noop {}
  {\bibfield  {journal} {\bibinfo  {journal} {Phys. Rev. A}\ }\textbf {\bibinfo
  {volume} {58}},\ \bibinfo {pages} {2733--2744} (\bibinfo {year}
  {1998})}\BibitemShut {NoStop}%
\bibitem [{\citenamefont {Vitali}\ and\ \citenamefont
  {Tombesi}(1999)}]{Vitali:99}%
  \BibitemOpen
  \bibfield  {author} {\bibinfo {author} {\bibfnamefont {D.}~\bibnamefont
  {Vitali}}\ and\ \bibinfo {author} {\bibfnamefont {P.}~\bibnamefont
  {Tombesi}},\ }\bibfield  {title} {\enquote {\bibinfo {title} {Using parity
  kicks for decoherence control},}\ }\href@noop {} {\bibfield  {journal}
  {\bibinfo  {journal} {Physical Review A}\ }\textbf {\bibinfo {volume} {59}},\
  \bibinfo {pages} {4178--4186} (\bibinfo {year} {1999})}\BibitemShut {NoStop}%
\bibitem [{\citenamefont {Zanardi}(1999)}]{Zan99}%
  \BibitemOpen
  \bibfield  {author} {\bibinfo {author} {\bibfnamefont {P.}~\bibnamefont
  {Zanardi}},\ }\bibfield  {title} {\enquote {\bibinfo {title} {Symmetrizing
  evolutions},}\ }\href {\doibase
  https://doi.org/10.1016/S0375-9601(99)00365-5} {\bibfield  {journal}
  {\bibinfo  {journal} {Physics Letters A}\ }\textbf {\bibinfo {volume}
  {258}},\ \bibinfo {pages} {77--82} (\bibinfo {year} {1999})}\BibitemShut
  {NoStop}%
\bibitem [{\citenamefont {Carr}\ and\ \citenamefont
  {Purcell}(1954)}]{carr1954effects}%
  \BibitemOpen
  \bibfield  {author} {\bibinfo {author} {\bibfnamefont {H.~Y.}\ \bibnamefont
  {Carr}}\ and\ \bibinfo {author} {\bibfnamefont {E.~M.}\ \bibnamefont
  {Purcell}},\ }\bibfield  {title} {\enquote {\bibinfo {title} {Effects of
  diffusion on free precession in nuclear magnetic resonance experiments},}\
  }\href {\doibase 10.1103/PhysRev.94.630} {\bibfield  {journal} {\bibinfo
  {journal} {Phys. Rev.}\ }\textbf {\bibinfo {volume} {94}},\ \bibinfo {pages}
  {630--638} (\bibinfo {year} {1954})}\BibitemShut {NoStop}%
\bibitem [{\citenamefont {Meiboom}\ and\ \citenamefont
  {Gill}(1958)}]{meiboom1958modified}%
  \BibitemOpen
  \bibfield  {author} {\bibinfo {author} {\bibfnamefont {Saul}\ \bibnamefont
  {Meiboom}}\ and\ \bibinfo {author} {\bibfnamefont {David}\ \bibnamefont
  {Gill}},\ }\bibfield  {title} {\enquote {\bibinfo {title} {Modified spin-echo
  method for measuring nuclear relaxation times},}\ }\href@noop {} {\bibfield
  {journal} {\bibinfo  {journal} {Review of scientific instruments}\ }\textbf
  {\bibinfo {volume} {29}},\ \bibinfo {pages} {688--691} (\bibinfo {year}
  {1958})}\BibitemShut {NoStop}%
\bibitem [{\citenamefont {Shor}(1995)}]{Sho95}%
  \BibitemOpen
  \bibfield  {author} {\bibinfo {author} {\bibfnamefont {P.~W.}\ \bibnamefont
  {Shor}},\ }\bibfield  {title} {\enquote {\bibinfo {title} {Scheme for
  reducing decoherence in quantum computer memory},}\ }\href {\doibase
  10.1103/PhysRevA.52.R2493} {\bibfield  {journal} {\bibinfo  {journal} {Phys.
  Rev. A}\ }\textbf {\bibinfo {volume} {52}},\ \bibinfo {pages} {R2493--R2496}
  (\bibinfo {year} {1995})}\BibitemShut {NoStop}%
\bibitem [{\citenamefont {Steane}(1996)}]{Ste96}%
  \BibitemOpen
  \bibfield  {author} {\bibinfo {author} {\bibfnamefont {A.~M.}\ \bibnamefont
  {Steane}},\ }\bibfield  {title} {\enquote {\bibinfo {title} {Error correcting
  codes in quantum theory},}\ }\href {\doibase 10.1103/PhysRevLett.77.793}
  {\bibfield  {journal} {\bibinfo  {journal} {Phys. Rev. Lett.}\ }\textbf
  {\bibinfo {volume} {77}},\ \bibinfo {pages} {793--797} (\bibinfo {year}
  {1996})}\BibitemShut {NoStop}%
\bibitem [{\citenamefont {Shor}(1996)}]{Sho96}%
  \BibitemOpen
  \bibfield  {author} {\bibinfo {author} {\bibfnamefont {P.}~\bibnamefont
  {Shor}},\ }\bibfield  {title} {\enquote {\bibinfo {title} {Fault-tolerant
  quantum computation},}\ }in\ \href {\doibase 10.1109/SFCS.1996.548464} {\emph
  {\bibinfo {booktitle} {Proceedings of 37th Conference on Foundations of
  Computer Science}}}\ (\bibinfo {year} {1996})\ pp.\ \bibinfo {pages}
  {56--65}\BibitemShut {NoStop}%
\bibitem [{\citenamefont {Preskill}(2018)}]{Pre18}%
  \BibitemOpen
  \bibfield  {author} {\bibinfo {author} {\bibfnamefont {J.}~\bibnamefont
  {Preskill}},\ }\bibfield  {title} {\enquote {\bibinfo {title} {Quantum
  {C}omputing in the {NISQ} era and beyond},}\ }\href {\doibase
  10.22331/q-2018-08-06-79} {\bibfield  {journal} {\bibinfo  {journal}
  {{Quantum}}\ }\textbf {\bibinfo {volume} {2}},\ \bibinfo {pages} {79}
  (\bibinfo {year} {2018})}\BibitemShut {NoStop}%
\bibitem [{\citenamefont {Ng}\ \emph {et~al.}(2011)\citenamefont {Ng},
  \citenamefont {Lidar},\ and\ \citenamefont {Preskill}}]{Ngh11}%
  \BibitemOpen
  \bibfield  {author} {\bibinfo {author} {\bibfnamefont {H.~K.}\ \bibnamefont
  {Ng}}, \bibinfo {author} {\bibfnamefont {D.~A.}\ \bibnamefont {Lidar}}, \
  and\ \bibinfo {author} {\bibfnamefont {J.}~\bibnamefont {Preskill}},\
  }\bibfield  {title} {\enquote {\bibinfo {title} {Combining dynamical
  decoupling with fault-tolerant quantum computation},}\ }\href {\doibase
  10.1103/PhysRevA.84.012305} {\bibfield  {journal} {\bibinfo  {journal} {Phys.
  Rev. A}\ }\textbf {\bibinfo {volume} {84}},\ \bibinfo {pages} {012305}
  (\bibinfo {year} {2011})}\BibitemShut {NoStop}%
\bibitem [{\citenamefont {Pokharel}\ \emph {et~al.}(2018)\citenamefont
  {Pokharel}, \citenamefont {Anand}, \citenamefont {Fortman},\ and\
  \citenamefont {Lidar}}]{pokharel2018demonstration}%
  \BibitemOpen
  \bibfield  {author} {\bibinfo {author} {\bibfnamefont {B.}~\bibnamefont
  {Pokharel}}, \bibinfo {author} {\bibfnamefont {N.}~\bibnamefont {Anand}},
  \bibinfo {author} {\bibfnamefont {B.}~\bibnamefont {Fortman}}, \ and\
  \bibinfo {author} {\bibfnamefont {D.~A.}\ \bibnamefont {Lidar}},\ }\bibfield
  {title} {\enquote {\bibinfo {title} {Demonstration of fidelity improvement
  using dynamical decoupling with superconducting qubits},}\ }\href@noop {}
  {\bibfield  {journal} {\bibinfo  {journal} {Physical Review Letters}\
  }\textbf {\bibinfo {volume} {121}},\ \bibinfo {pages} {220502} (\bibinfo
  {year} {2018})}\BibitemShut {NoStop}%
\bibitem [{\citenamefont {Ezzell}\ \emph {et~al.}(2023)\citenamefont {Ezzell},
  \citenamefont {Pokharel}, \citenamefont {Tewala}, \citenamefont {Quiroz},\
  and\ \citenamefont {Lidar}}]{Ezz22}%
  \BibitemOpen
  \bibfield  {author} {\bibinfo {author} {\bibfnamefont {Nic}\ \bibnamefont
  {Ezzell}}, \bibinfo {author} {\bibfnamefont {Bibek}\ \bibnamefont
  {Pokharel}}, \bibinfo {author} {\bibfnamefont {Lina}\ \bibnamefont {Tewala}},
  \bibinfo {author} {\bibfnamefont {Gregory}\ \bibnamefont {Quiroz}}, \ and\
  \bibinfo {author} {\bibfnamefont {Daniel~A.}\ \bibnamefont {Lidar}},\
  }\bibfield  {title} {\enquote {\bibinfo {title} {Dynamical decoupling for
  superconducting qubits: A performance survey},}\ }\href {\doibase
  10.1103/PhysRevApplied.20.064027} {\bibfield  {journal} {\bibinfo  {journal}
  {Phys. Rev. Appl.}\ }\textbf {\bibinfo {volume} {20}},\ \bibinfo {pages}
  {064027} (\bibinfo {year} {2023})}\BibitemShut {NoStop}%
\bibitem [{\citenamefont {Jurcevic}\ \emph {et~al.}(2021)\citenamefont
  {Jurcevic}, \citenamefont {Javadi-Abhari}, \citenamefont {Bishop},
  \citenamefont {Lauer}, \citenamefont {Bogorin}, \citenamefont {Brink},
  \citenamefont {Capelluto}, \citenamefont {Günlük}, \citenamefont {Itoko},
  \citenamefont {Kanazawa}, \citenamefont {Kandala}, \citenamefont {Keefe},
  \citenamefont {Krsulich}, \citenamefont {Landers}, \citenamefont
  {Lewandowski}, \citenamefont {McClure}, \citenamefont {Nannicini},
  \citenamefont {Narasgond}, \citenamefont {Nayfeh}, \citenamefont {Pritchett},
  \citenamefont {Rothwell}, \citenamefont {Srinivasan}, \citenamefont
  {Sundaresan}, \citenamefont {Wang}, \citenamefont {Wei}, \citenamefont
  {Wood}, \citenamefont {Yau}, \citenamefont {Zhang}, \citenamefont {Dial},
  \citenamefont {Chow},\ and\ \citenamefont {Gambetta}}]{Jurcevic_2021}%
  \BibitemOpen
  \bibfield  {author} {\bibinfo {author} {\bibfnamefont {Petar}\ \bibnamefont
  {Jurcevic}}, \bibinfo {author} {\bibfnamefont {Ali}\ \bibnamefont
  {Javadi-Abhari}}, \bibinfo {author} {\bibfnamefont {Lev~S}\ \bibnamefont
  {Bishop}}, \bibinfo {author} {\bibfnamefont {Isaac}\ \bibnamefont {Lauer}},
  \bibinfo {author} {\bibfnamefont {Daniela~F}\ \bibnamefont {Bogorin}},
  \bibinfo {author} {\bibfnamefont {Markus}\ \bibnamefont {Brink}}, \bibinfo
  {author} {\bibfnamefont {Lauren}\ \bibnamefont {Capelluto}}, \bibinfo
  {author} {\bibfnamefont {Oktay}\ \bibnamefont {Günlük}}, \bibinfo {author}
  {\bibfnamefont {Toshinari}\ \bibnamefont {Itoko}}, \bibinfo {author}
  {\bibfnamefont {Naoki}\ \bibnamefont {Kanazawa}}, \bibinfo {author}
  {\bibfnamefont {Abhinav}\ \bibnamefont {Kandala}}, \bibinfo {author}
  {\bibfnamefont {George~A}\ \bibnamefont {Keefe}}, \bibinfo {author}
  {\bibfnamefont {Kevin}\ \bibnamefont {Krsulich}}, \bibinfo {author}
  {\bibfnamefont {William}\ \bibnamefont {Landers}}, \bibinfo {author}
  {\bibfnamefont {Eric~P}\ \bibnamefont {Lewandowski}}, \bibinfo {author}
  {\bibfnamefont {Douglas~T}\ \bibnamefont {McClure}}, \bibinfo {author}
  {\bibfnamefont {Giacomo}\ \bibnamefont {Nannicini}}, \bibinfo {author}
  {\bibfnamefont {Adinath}\ \bibnamefont {Narasgond}}, \bibinfo {author}
  {\bibfnamefont {Hasan~M}\ \bibnamefont {Nayfeh}}, \bibinfo {author}
  {\bibfnamefont {Emily}\ \bibnamefont {Pritchett}}, \bibinfo {author}
  {\bibfnamefont {Mary~Beth}\ \bibnamefont {Rothwell}}, \bibinfo {author}
  {\bibfnamefont {Srikanth}\ \bibnamefont {Srinivasan}}, \bibinfo {author}
  {\bibfnamefont {Neereja}\ \bibnamefont {Sundaresan}}, \bibinfo {author}
  {\bibfnamefont {Cindy}\ \bibnamefont {Wang}}, \bibinfo {author}
  {\bibfnamefont {Ken~X}\ \bibnamefont {Wei}}, \bibinfo {author} {\bibfnamefont
  {Christopher~J}\ \bibnamefont {Wood}}, \bibinfo {author} {\bibfnamefont
  {Jeng-Bang}\ \bibnamefont {Yau}}, \bibinfo {author} {\bibfnamefont {Eric~J}\
  \bibnamefont {Zhang}}, \bibinfo {author} {\bibfnamefont {Oliver~E}\
  \bibnamefont {Dial}}, \bibinfo {author} {\bibfnamefont {Jerry~M}\
  \bibnamefont {Chow}}, \ and\ \bibinfo {author} {\bibfnamefont {Jay~M}\
  \bibnamefont {Gambetta}},\ }\bibfield  {title} {\enquote {\bibinfo {title}
  {Demonstration of quantum volume 64 on a superconducting quantum computing
  system},}\ }\href {\doibase 10.1088/2058-9565/abe519} {\bibfield  {journal}
  {\bibinfo  {journal} {Quantum Science and Technology}\ }\textbf {\bibinfo
  {volume} {6}},\ \bibinfo {pages} {025020} (\bibinfo {year}
  {2021})}\BibitemShut {NoStop}%
\bibitem [{\citenamefont {AI}(2023)}]{Goo23}%
  \BibitemOpen
  \bibfield  {author} {\bibinfo {author} {\bibfnamefont {G.~Q.}\ \bibnamefont
  {AI}},\ }\bibfield  {title} {\enquote {\bibinfo {title} {Suppressing quantum
  errors by scaling a surface code logical qubit},}\ }\href@noop {} {\bibfield
  {journal} {\bibinfo  {journal} {Nature}\ }\textbf {\bibinfo {volume} {614}},\
  \bibinfo {pages} {676--681} (\bibinfo {year} {2023})}\BibitemShut {NoStop}%
\bibitem [{\citenamefont {Kim}\ \emph {et~al.}(2023{\natexlab{a}})\citenamefont
  {Kim}, \citenamefont {Wood}, \citenamefont {Yoder}, \citenamefont {Merkel},
  \citenamefont {Gambetta}, \citenamefont {Temme},\ and\ \citenamefont
  {Kandala}}]{Kim23}%
  \BibitemOpen
  \bibfield  {author} {\bibinfo {author} {\bibfnamefont {Y.}~\bibnamefont
  {Kim}}, \bibinfo {author} {\bibfnamefont {C.~J.}\ \bibnamefont {Wood}},
  \bibinfo {author} {\bibfnamefont {T.~J.}\ \bibnamefont {Yoder}}, \bibinfo
  {author} {\bibfnamefont {S.~T.}\ \bibnamefont {Merkel}}, \bibinfo {author}
  {\bibfnamefont {J.~M.}\ \bibnamefont {Gambetta}}, \bibinfo {author}
  {\bibfnamefont {K.}~\bibnamefont {Temme}}, \ and\ \bibinfo {author}
  {\bibfnamefont {A.}~\bibnamefont {Kandala}},\ }\bibfield  {title} {\enquote
  {\bibinfo {title} {Scalable error mitigation for noisy quantum circuits
  produces competitive expectation values},}\ }\href@noop {} {\bibfield
  {journal} {\bibinfo  {journal} {Nat. Phys.}\ }\textbf {\bibinfo {volume}
  {19}},\ \bibinfo {pages} {752--759} (\bibinfo {year}
  {2023}{\natexlab{a}})}\BibitemShut {NoStop}%
\bibitem [{\citenamefont {Kim}\ \emph {et~al.}(2023{\natexlab{b}})\citenamefont
  {Kim}, \citenamefont {Eddins}, \citenamefont {Anand}, \citenamefont {Wei},
  \citenamefont {Van Den~Berg}, \citenamefont {Rosenblatt}, \citenamefont
  {Nayfeh}, \citenamefont {Wu}, \citenamefont {Zaletel}, \citenamefont
  {Temme},\ and\ \citenamefont {Kandala}}]{Kim23_2}%
  \BibitemOpen
  \bibfield  {author} {\bibinfo {author} {\bibfnamefont {Y.}~\bibnamefont
  {Kim}}, \bibinfo {author} {\bibfnamefont {A.}~\bibnamefont {Eddins}},
  \bibinfo {author} {\bibfnamefont {S.}~\bibnamefont {Anand}}, \bibinfo
  {author} {\bibfnamefont {K.}~\bibnamefont {Wei}}, \bibinfo {author}
  {\bibfnamefont {E.}~\bibnamefont {Van Den~Berg}}, \bibinfo {author}
  {\bibfnamefont {S.}~\bibnamefont {Rosenblatt}}, \bibinfo {author}
  {\bibfnamefont {H.}~\bibnamefont {Nayfeh}}, \bibinfo {author} {\bibfnamefont
  {Y.}~\bibnamefont {Wu}}, \bibinfo {author} {\bibfnamefont {M.}~\bibnamefont
  {Zaletel}}, \bibinfo {author} {\bibfnamefont {K.}~\bibnamefont {Temme}}, \
  and\ \bibinfo {author} {\bibfnamefont {A.}~\bibnamefont {Kandala}},\
  }\bibfield  {title} {\enquote {\bibinfo {title} {Evidence for the utility of
  quantum computing before fault tolerance},}\ }\href@noop {} {\bibfield
  {journal} {\bibinfo  {journal} {Nature}\ }\textbf {\bibinfo {volume} {618}},\
  \bibinfo {pages} {500--505} (\bibinfo {year}
  {2023}{\natexlab{b}})}\BibitemShut {NoStop}%
\bibitem [{\citenamefont {Biercuk}\ \emph {et~al.}(2009)\citenamefont
  {Biercuk}, \citenamefont {Uys}, \citenamefont {VanDevender}, \citenamefont
  {Shiga}, \citenamefont {Itano},\ and\ \citenamefont {Bollinger}}]{Bie09}%
  \BibitemOpen
  \bibfield  {author} {\bibinfo {author} {\bibfnamefont {M.~J.}\ \bibnamefont
  {Biercuk}}, \bibinfo {author} {\bibfnamefont {H.}~\bibnamefont {Uys}},
  \bibinfo {author} {\bibfnamefont {A.~P.}\ \bibnamefont {VanDevender}},
  \bibinfo {author} {\bibfnamefont {N.}~\bibnamefont {Shiga}}, \bibinfo
  {author} {\bibfnamefont {W.~M.}\ \bibnamefont {Itano}}, \ and\ \bibinfo
  {author} {\bibfnamefont {J.~J.}\ \bibnamefont {Bollinger}},\ }\bibfield
  {title} {\enquote {\bibinfo {title} {Experimental {U}hrig dynamical
  decoupling using trapped ions},}\ }\href {\doibase
  10.1103/PhysRevA.79.062324} {\bibfield  {journal} {\bibinfo  {journal} {Phys.
  Rev. A}\ }\textbf {\bibinfo {volume} {79}},\ \bibinfo {pages} {062324}
  (\bibinfo {year} {2009})}\BibitemShut {NoStop}%
\bibitem [{\citenamefont {Du}\ \emph {et~al.}(2009)\citenamefont {Du},
  \citenamefont {Rong}, \citenamefont {Zhao}, \citenamefont {Wang},
  \citenamefont {Yang},\ and\ \citenamefont {Liu}}]{Duj09}%
  \BibitemOpen
  \bibfield  {author} {\bibinfo {author} {\bibfnamefont {J.}~\bibnamefont
  {Du}}, \bibinfo {author} {\bibfnamefont {X.}~\bibnamefont {Rong}}, \bibinfo
  {author} {\bibfnamefont {N.}~\bibnamefont {Zhao}}, \bibinfo {author}
  {\bibfnamefont {Y.}~\bibnamefont {Wang}}, \bibinfo {author} {\bibfnamefont
  {J.}~\bibnamefont {Yang}}, \ and\ \bibinfo {author} {\bibfnamefont {R.~B.}\
  \bibnamefont {Liu}},\ }\bibfield  {title} {\enquote {\bibinfo {title}
  {Preserving electron spin coherence in solids by optimal dynamical
  decoupling},}\ }\href@noop {} {\bibfield  {journal} {\bibinfo  {journal}
  {Nature}\ }\textbf {\bibinfo {volume} {461}},\ \bibinfo {pages} {1265--1268}
  (\bibinfo {year} {2009})}\BibitemShut {NoStop}%
\bibitem [{\citenamefont {de~Lange}\ \emph {et~al.}(2010)\citenamefont
  {de~Lange}, \citenamefont {Wang}, \citenamefont {Riste}, \citenamefont
  {Dobrovitski},\ and\ \citenamefont {Hanson}}]{de2010universal}%
  \BibitemOpen
  \bibfield  {author} {\bibinfo {author} {\bibfnamefont {G}~\bibnamefont
  {de~Lange}}, \bibinfo {author} {\bibfnamefont {ZH}~\bibnamefont {Wang}},
  \bibinfo {author} {\bibfnamefont {D}~\bibnamefont {Riste}}, \bibinfo {author}
  {\bibfnamefont {VV}~\bibnamefont {Dobrovitski}}, \ and\ \bibinfo {author}
  {\bibfnamefont {R}~\bibnamefont {Hanson}},\ }\bibfield  {title} {\enquote
  {\bibinfo {title} {Universal dynamical decoupling of a single solid-state
  spin from a spin bath},}\ }\href@noop {} {\bibfield  {journal} {\bibinfo
  {journal} {Science}\ }\textbf {\bibinfo {volume} {330}},\ \bibinfo {pages}
  {60--63} (\bibinfo {year} {2010})}\BibitemShut {NoStop}%
\bibitem [{\citenamefont {Wang}\ \emph {et~al.}(2012)\citenamefont {Wang},
  \citenamefont {de~Lange}, \citenamefont {Rist\`e}, \citenamefont {Hanson},\
  and\ \citenamefont {Dobrovitski}}]{Wan12}%
  \BibitemOpen
  \bibfield  {author} {\bibinfo {author} {\bibfnamefont {Z.}~\bibnamefont
  {Wang}}, \bibinfo {author} {\bibfnamefont {G.}~\bibnamefont {de~Lange}},
  \bibinfo {author} {\bibfnamefont {D.}~\bibnamefont {Rist\`e}}, \bibinfo
  {author} {\bibfnamefont {R.}~\bibnamefont {Hanson}}, \ and\ \bibinfo {author}
  {\bibfnamefont {V.~V.}\ \bibnamefont {Dobrovitski}},\ }\bibfield  {title}
  {\enquote {\bibinfo {title} {Comparison of dynamical decoupling protocols for
  a nitrogen-vacancy center in diamond},}\ }\href@noop {} {\bibfield  {journal}
  {\bibinfo  {journal} {Phys. Rev. B}\ }\textbf {\bibinfo {volume} {85}},\
  \bibinfo {pages} {155204} (\bibinfo {year} {2012})}\BibitemShut {NoStop}%
\bibitem [{\citenamefont {Farfurnik}\ \emph {et~al.}(2016)\citenamefont
  {Farfurnik}, \citenamefont {Jarmola}, \citenamefont {Pham}, \citenamefont
  {Wang}, \citenamefont {Dobrovitski}, \citenamefont {Walsworth}, \citenamefont
  {Budker},\ and\ \citenamefont {Bar-Gill}}]{farfurnik2016improving}%
  \BibitemOpen
  \bibfield  {author} {\bibinfo {author} {\bibfnamefont {D}~\bibnamefont
  {Farfurnik}}, \bibinfo {author} {\bibfnamefont {A}~\bibnamefont {Jarmola}},
  \bibinfo {author} {\bibfnamefont {LM}~\bibnamefont {Pham}}, \bibinfo {author}
  {\bibfnamefont {ZH}~\bibnamefont {Wang}}, \bibinfo {author} {\bibfnamefont
  {VV}~\bibnamefont {Dobrovitski}}, \bibinfo {author} {\bibfnamefont
  {RL}~\bibnamefont {Walsworth}}, \bibinfo {author} {\bibfnamefont
  {D}~\bibnamefont {Budker}}, \ and\ \bibinfo {author} {\bibfnamefont
  {N}~\bibnamefont {Bar-Gill}},\ }\bibfield  {title} {\enquote {\bibinfo
  {title} {Improving the coherence properties of solid-state spin ensembles via
  optimized dynamical decoupling},}\ }in\ \href@noop {} {\emph {\bibinfo
  {booktitle} {Quantum Optics}}},\ Vol.\ \bibinfo {volume} {9900}\ (\bibinfo
  {organization} {SPIE},\ \bibinfo {year} {2016})\ pp.\ \bibinfo {pages}
  {111--120}\BibitemShut {NoStop}%
\bibitem [{\citenamefont {Genov}\ \emph {et~al.}(2017)\citenamefont {Genov},
  \citenamefont {Schraft}, \citenamefont {Vitanov},\ and\ \citenamefont
  {Halfmann}}]{Gen17}%
  \BibitemOpen
  \bibfield  {author} {\bibinfo {author} {\bibfnamefont {G.~T.}\ \bibnamefont
  {Genov}}, \bibinfo {author} {\bibfnamefont {D.}~\bibnamefont {Schraft}},
  \bibinfo {author} {\bibfnamefont {N.~V.}\ \bibnamefont {Vitanov}}, \ and\
  \bibinfo {author} {\bibfnamefont {T.}~\bibnamefont {Halfmann}},\ }\bibfield
  {title} {\enquote {\bibinfo {title} {Arbitrarily accurate pulse sequences for
  robust dynamical decoupling},}\ }\href@noop {} {\bibfield  {journal}
  {\bibinfo  {journal} {Phys. Rev. Lett.}\ }\textbf {\bibinfo {volume} {118}},\
  \bibinfo {pages} {133202} (\bibinfo {year} {2017})}\BibitemShut {NoStop}%
\bibitem [{\citenamefont {Viola}\ and\ \citenamefont {Knill}(2009)}]{Vio03}%
  \BibitemOpen
  \bibfield  {author} {\bibinfo {author} {\bibfnamefont {L.}~\bibnamefont
  {Viola}}\ and\ \bibinfo {author} {\bibfnamefont {E.}~\bibnamefont {Knill}},\
  }\bibfield  {title} {\enquote {\bibinfo {title} {Robust dynamical decoupling
  of quantum systems with bounded controls},}\ }\href@noop {} {\bibfield
  {journal} {\bibinfo  {journal} {Phys. Rev. Lett.}\ }\textbf {\bibinfo
  {volume} {90}},\ \bibinfo {pages} {037901} (\bibinfo {year}
  {2009})}\BibitemShut {NoStop}%
\bibitem [{\citenamefont {Singkanipa}\ \emph {et~al.}(2024)\citenamefont
  {Singkanipa}, \citenamefont {Kasatkin}, \citenamefont {Zhou}, \citenamefont
  {Quiroz},\ and\ \citenamefont {Lidar}}]{singkanipa2024demonstration}%
  \BibitemOpen
  \bibfield  {author} {\bibinfo {author} {\bibfnamefont {P.}~\bibnamefont
  {Singkanipa}}, \bibinfo {author} {\bibfnamefont {V.}~\bibnamefont
  {Kasatkin}}, \bibinfo {author} {\bibfnamefont {Z.}~\bibnamefont {Zhou}},
  \bibinfo {author} {\bibfnamefont {G.}~\bibnamefont {Quiroz}}, \ and\ \bibinfo
  {author} {\bibfnamefont {D.~A.}\ \bibnamefont {Lidar}},\ }\bibfield  {title}
  {\enquote {\bibinfo {title} {Demonstration of algorithmic quantum speedup for
  an abelian hidden subgroup problem},}\ }\href@noop {} {\bibfield  {journal}
  {\bibinfo  {journal} {arXiv preprint arXiv:2401.07934}\ } (\bibinfo {year}
  {2024})}\BibitemShut {NoStop}%
\bibitem [{\citenamefont {Pokharel}\ and\ \citenamefont {Lidar}(2024)}]{Pok22}%
  \BibitemOpen
  \bibfield  {author} {\bibinfo {author} {\bibfnamefont {B.}~\bibnamefont
  {Pokharel}}\ and\ \bibinfo {author} {\bibfnamefont {D.A.}\ \bibnamefont
  {Lidar}},\ }\bibfield  {title} {\enquote {\bibinfo {title}
  {Better-than-classical grover search via quantum error detection and
  suppression},}\ }\href {\doibase https://doi.org/10.1038/s41534-023-00794-6}
  {\bibfield  {journal} {\bibinfo  {journal} {npj Quantum Inf}\ }\textbf
  {\bibinfo {volume} {10}} (\bibinfo {year} {2024}),\
  https://doi.org/10.1038/s41534-023-00794-6}\BibitemShut {NoStop}%
\bibitem [{\citenamefont {Pokharel}\ and\ \citenamefont {Lidar}(2023)}]{Pok23}%
  \BibitemOpen
  \bibfield  {author} {\bibinfo {author} {\bibfnamefont {B.}~\bibnamefont
  {Pokharel}}\ and\ \bibinfo {author} {\bibfnamefont {D.~A.}\ \bibnamefont
  {Lidar}},\ }\bibfield  {title} {\enquote {\bibinfo {title} {Demonstration of
  algorithmic quantum speedup},}\ }\href {\doibase
  10.1103/PhysRevLett.130.210602} {\bibfield  {journal} {\bibinfo  {journal}
  {Phys. Rev. Lett.}\ }\textbf {\bibinfo {volume} {130}},\ \bibinfo {pages}
  {210602} (\bibinfo {year} {2023})}\BibitemShut {NoStop}%
\bibitem [{\citenamefont {Tripathi}\ \emph {et~al.}(2022)\citenamefont
  {Tripathi}, \citenamefont {Chen}, \citenamefont {Khezri}, \citenamefont
  {Yip}, \citenamefont {Levenson-Falk},\ and\ \citenamefont {Lidar}}]{Tri22}%
  \BibitemOpen
  \bibfield  {author} {\bibinfo {author} {\bibfnamefont {V.}~\bibnamefont
  {Tripathi}}, \bibinfo {author} {\bibfnamefont {H.}~\bibnamefont {Chen}},
  \bibinfo {author} {\bibfnamefont {M.}~\bibnamefont {Khezri}}, \bibinfo
  {author} {\bibfnamefont {K.}~\bibnamefont {Yip}}, \bibinfo {author}
  {\bibfnamefont {E.}~\bibnamefont {Levenson-Falk}}, \ and\ \bibinfo {author}
  {\bibfnamefont {D.~A.}\ \bibnamefont {Lidar}},\ }\bibfield  {title} {\enquote
  {\bibinfo {title} {Suppression of crosstalk in superconducting qubits using
  dynamical decoupling},}\ }\href {\doibase 10.1103/PhysRevApplied.18.024068}
  {\bibfield  {journal} {\bibinfo  {journal} {Phys. Rev. Appl.}\ }\textbf
  {\bibinfo {volume} {18}},\ \bibinfo {pages} {024068} (\bibinfo {year}
  {2022})}\BibitemShut {NoStop}%
\bibitem [{\citenamefont {Zhou}\ \emph {et~al.}(2023)\citenamefont {Zhou},
  \citenamefont {Sitler}, \citenamefont {Oda}, \citenamefont {Schultz},\ and\
  \citenamefont {Quiroz}}]{Zho23}%
  \BibitemOpen
  \bibfield  {author} {\bibinfo {author} {\bibfnamefont {Z.}~\bibnamefont
  {Zhou}}, \bibinfo {author} {\bibfnamefont {R.}~\bibnamefont {Sitler}},
  \bibinfo {author} {\bibfnamefont {Y.}~\bibnamefont {Oda}}, \bibinfo {author}
  {\bibfnamefont {K.}~\bibnamefont {Schultz}}, \ and\ \bibinfo {author}
  {\bibfnamefont {G.}~\bibnamefont {Quiroz}},\ }\bibfield  {title} {\enquote
  {\bibinfo {title} {Quantum crosstalk robust quantum control},}\ }\href@noop
  {} {\bibfield  {journal} {\bibinfo  {journal} {Phys. Rev. Lett.}\ }\textbf
  {\bibinfo {volume} {131}},\ \bibinfo {pages} {210802} (\bibinfo {year}
  {2023})}\BibitemShut {NoStop}%
\bibitem [{\citenamefont {Shirizly}\ \emph {et~al.}(2024)\citenamefont
  {Shirizly}, \citenamefont {Misguich},\ and\ \citenamefont {Landa}}]{Shi24}%
  \BibitemOpen
  \bibfield  {author} {\bibinfo {author} {\bibfnamefont {Liran}\ \bibnamefont
  {Shirizly}}, \bibinfo {author} {\bibfnamefont {Gr\'egoire}\ \bibnamefont
  {Misguich}}, \ and\ \bibinfo {author} {\bibfnamefont {Haggai}\ \bibnamefont
  {Landa}},\ }\bibfield  {title} {\enquote {\bibinfo {title} {Dissipative
  dynamics of graph-state stabilizers with superconducting qubits},}\ }\href
  {\doibase 10.1103/PhysRevLett.132.010601} {\bibfield  {journal} {\bibinfo
  {journal} {Phys. Rev. Lett.}\ }\textbf {\bibinfo {volume} {132}},\ \bibinfo
  {pages} {010601} (\bibinfo {year} {2024})}\BibitemShut {NoStop}%
\bibitem [{\citenamefont {Quiroz}\ \emph {et~al.}(2024)\citenamefont {Quiroz},
  \citenamefont {Pokharel}, \citenamefont {Boen}, \citenamefont {Tewala},
  \citenamefont {Tripathi}, \citenamefont {Williams}, \citenamefont {Wu},
  \citenamefont {Titum}, \citenamefont {Schultz},\ and\ \citenamefont
  {Lidar}}]{Qui24}%
  \BibitemOpen
  \bibfield  {author} {\bibinfo {author} {\bibfnamefont {Gregory}\ \bibnamefont
  {Quiroz}}, \bibinfo {author} {\bibfnamefont {Bibek}\ \bibnamefont
  {Pokharel}}, \bibinfo {author} {\bibfnamefont {Joseph}\ \bibnamefont {Boen}},
  \bibinfo {author} {\bibfnamefont {Lina}\ \bibnamefont {Tewala}}, \bibinfo
  {author} {\bibfnamefont {Vinay}\ \bibnamefont {Tripathi}}, \bibinfo {author}
  {\bibfnamefont {Devon}\ \bibnamefont {Williams}}, \bibinfo {author}
  {\bibfnamefont {Lian-Ao}\ \bibnamefont {Wu}}, \bibinfo {author}
  {\bibfnamefont {Paraj}\ \bibnamefont {Titum}}, \bibinfo {author}
  {\bibfnamefont {Kevin}\ \bibnamefont {Schultz}}, \ and\ \bibinfo {author}
  {\bibfnamefont {Daniel}\ \bibnamefont {Lidar}},\ }\bibfield  {title}
  {\enquote {\bibinfo {title} {Dynamically generated decoherence-free subspaces
  and subsystems on superconducting qubits},}\ }\href {\doibase
  10.1088/1361-6633/ad6805} {\bibfield  {journal} {\bibinfo  {journal} {Rep.
  Prog. Phys.}\ }\textbf {\bibinfo {volume} {87}},\ \bibinfo {pages} {097601}
  (\bibinfo {year} {2024})}\BibitemShut {NoStop}%
\bibitem [{\citenamefont {Krantz}\ \emph {et~al.}(2019)\citenamefont {Krantz},
  \citenamefont {Kjaergaard}, \citenamefont {Yan}, \citenamefont {Orlando},
  \citenamefont {Gustavsson},\ and\ \citenamefont
  {Oliver}}]{krantz2019quantum}%
  \BibitemOpen
  \bibfield  {author} {\bibinfo {author} {\bibfnamefont {Philip}\ \bibnamefont
  {Krantz}}, \bibinfo {author} {\bibfnamefont {Morten}\ \bibnamefont
  {Kjaergaard}}, \bibinfo {author} {\bibfnamefont {Fei}\ \bibnamefont {Yan}},
  \bibinfo {author} {\bibfnamefont {Terry~P}\ \bibnamefont {Orlando}}, \bibinfo
  {author} {\bibfnamefont {Simon}\ \bibnamefont {Gustavsson}}, \ and\ \bibinfo
  {author} {\bibfnamefont {William~D}\ \bibnamefont {Oliver}},\ }\bibfield
  {title} {\enquote {\bibinfo {title} {A quantum engineer's guide to
  superconducting qubits},}\ }\href@noop {} {\bibfield  {journal} {\bibinfo
  {journal} {Applied physics reviews}\ }\textbf {\bibinfo {volume} {6}}
  (\bibinfo {year} {2019})}\BibitemShut {NoStop}%
\bibitem [{\citenamefont {Zhang}\ \emph {et~al.}(2022)\citenamefont {Zhang},
  \citenamefont {Pokharel}, \citenamefont {Levenson-Falk},\ and\ \citenamefont
  {Lidar}}]{zhang2022predicting}%
  \BibitemOpen
  \bibfield  {author} {\bibinfo {author} {\bibfnamefont {Haimeng}\ \bibnamefont
  {Zhang}}, \bibinfo {author} {\bibfnamefont {Bibek}\ \bibnamefont {Pokharel}},
  \bibinfo {author} {\bibfnamefont {EM}~\bibnamefont {Levenson-Falk}}, \ and\
  \bibinfo {author} {\bibfnamefont {Daniel}\ \bibnamefont {Lidar}},\ }\bibfield
   {title} {\enquote {\bibinfo {title} {Predicting non-markovian
  superconducting-qubit dynamics from tomographic reconstruction},}\
  }\href@noop {} {\bibfield  {journal} {\bibinfo  {journal} {Physical Review
  Applied}\ }\textbf {\bibinfo {volume} {17}},\ \bibinfo {pages} {054018}
  (\bibinfo {year} {2022})}\BibitemShut {NoStop}%
\bibitem [{\citenamefont {C{\'o}rcoles}\ \emph {et~al.}(2019)\citenamefont
  {C{\'o}rcoles}, \citenamefont {Mezzacapo}, \citenamefont {Chow},\ and\
  \citenamefont {Gambetta}}]{kandala2019error}%
  \BibitemOpen
  \bibfield  {author} {\bibinfo {author} {\bibfnamefont {A.~D.}\ \bibnamefont
  {C{\'o}rcoles}}, \bibinfo {author} {\bibfnamefont {A.}~\bibnamefont
  {Mezzacapo}}, \bibinfo {author} {\bibfnamefont {J.~M.}\ \bibnamefont {Chow}},
  \ and\ \bibinfo {author} {\bibfnamefont {J.~M.}\ \bibnamefont {Gambetta}},\
  }\bibfield  {title} {\enquote {\bibinfo {title} {Error mitigation extends the
  computational reach of a noisy quantum processor},}\ }\href@noop {}
  {\bibfield  {journal} {\bibinfo  {journal} {Nature}\ }\textbf {\bibinfo
  {volume} {567}},\ \bibinfo {pages} {491--495} (\bibinfo {year}
  {2019})}\BibitemShut {NoStop}%
\bibitem [{\citenamefont {Van Den~Berg}\ \emph {et~al.}(2023)\citenamefont {Van
  Den~Berg}, \citenamefont {Minev}, \citenamefont {Kandala},\ and\
  \citenamefont {Temme}}]{van2023probabilistic}%
  \BibitemOpen
  \bibfield  {author} {\bibinfo {author} {\bibfnamefont {E.}~\bibnamefont {Van
  Den~Berg}}, \bibinfo {author} {\bibfnamefont {Z.~K.}\ \bibnamefont {Minev}},
  \bibinfo {author} {\bibfnamefont {A.}~\bibnamefont {Kandala}}, \ and\
  \bibinfo {author} {\bibfnamefont {K.}~\bibnamefont {Temme}},\ }\bibfield
  {title} {\enquote {\bibinfo {title} {Probabilistic error cancellation with
  sparse pauli--lindblad models on noisy quantum processors},}\ }\href@noop {}
  {\bibfield  {journal} {\bibinfo  {journal} {Nature Physics}\ ,\ \bibinfo
  {pages} {1--6}} (\bibinfo {year} {2023})}\BibitemShut {NoStop}%
\bibitem [{\citenamefont {Temme}\ \emph {et~al.}(2017)\citenamefont {Temme},
  \citenamefont {Bravyi},\ and\ \citenamefont {Gambetta}}]{temme2017error}%
  \BibitemOpen
  \bibfield  {author} {\bibinfo {author} {\bibfnamefont {K.}~\bibnamefont
  {Temme}}, \bibinfo {author} {\bibfnamefont {S.}~\bibnamefont {Bravyi}}, \
  and\ \bibinfo {author} {\bibfnamefont {J.~M.}\ \bibnamefont {Gambetta}},\
  }\bibfield  {title} {\enquote {\bibinfo {title} {Error mitigation for
  short-depth quantum circuits},}\ }\href@noop {} {\bibfield  {journal}
  {\bibinfo  {journal} {Physical Review Letters}\ }\textbf {\bibinfo {volume}
  {119}},\ \bibinfo {pages} {180509} (\bibinfo {year} {2017})}\BibitemShut
  {NoStop}%
\bibitem [{\citenamefont {Nation}\ \emph {et~al.}(2021)\citenamefont {Nation},
  \citenamefont {Kang}, \citenamefont {Sundaresan},\ and\ \citenamefont
  {Gambetta}}]{nation2021scalable}%
  \BibitemOpen
  \bibfield  {author} {\bibinfo {author} {\bibfnamefont {P.~D.}\ \bibnamefont
  {Nation}}, \bibinfo {author} {\bibfnamefont {H.}~\bibnamefont {Kang}},
  \bibinfo {author} {\bibfnamefont {N.}~\bibnamefont {Sundaresan}}, \ and\
  \bibinfo {author} {\bibfnamefont {J.~M.}\ \bibnamefont {Gambetta}},\
  }\bibfield  {title} {\enquote {\bibinfo {title} {Scalable mitigation of
  measurement errors on quantum computers},}\ }\href@noop {} {\bibfield
  {journal} {\bibinfo  {journal} {PRX Quantum}\ }\textbf {\bibinfo {volume}
  {2}},\ \bibinfo {pages} {040326} (\bibinfo {year} {2021})}\BibitemShut
  {NoStop}%
\bibitem [{\citenamefont {Van Den~Berg}\ \emph {et~al.}(2022)\citenamefont {Van
  Den~Berg}, \citenamefont {Minev},\ and\ \citenamefont
  {Temme}}]{van2022model}%
  \BibitemOpen
  \bibfield  {author} {\bibinfo {author} {\bibfnamefont {E.}~\bibnamefont {Van
  Den~Berg}}, \bibinfo {author} {\bibfnamefont {Z.~K.}\ \bibnamefont {Minev}},
  \ and\ \bibinfo {author} {\bibfnamefont {K.}~\bibnamefont {Temme}},\
  }\bibfield  {title} {\enquote {\bibinfo {title} {Model-free readout-error
  mitigation for quantum expectation values},}\ }\href@noop {} {\bibfield
  {journal} {\bibinfo  {journal} {Physical Review A}\ }\textbf {\bibinfo
  {volume} {105}},\ \bibinfo {pages} {032620} (\bibinfo {year}
  {2022})}\BibitemShut {NoStop}%
\bibitem [{\citenamefont {Pokharel}\ \emph {et~al.}(2024)\citenamefont
  {Pokharel}, \citenamefont {Srinivasan}, \citenamefont {Quiroz},\ and\
  \citenamefont {Boots}}]{pokharel2024scalable}%
  \BibitemOpen
  \bibfield  {author} {\bibinfo {author} {\bibfnamefont {B.}~\bibnamefont
  {Pokharel}}, \bibinfo {author} {\bibfnamefont {S.}~\bibnamefont
  {Srinivasan}}, \bibinfo {author} {\bibfnamefont {G.}~\bibnamefont {Quiroz}},
  \ and\ \bibinfo {author} {\bibfnamefont {B.}~\bibnamefont {Boots}},\
  }\bibfield  {title} {\enquote {\bibinfo {title} {Scalable measurement error
  mitigation via iterative bayesian unfolding},}\ }\href@noop {} {\bibfield
  {journal} {\bibinfo  {journal} {Physical Review Research}\ }\textbf {\bibinfo
  {volume} {6}},\ \bibinfo {pages} {013187} (\bibinfo {year}
  {2024})}\BibitemShut {NoStop}%
\bibitem [{\citenamefont {Yang}\ \emph {et~al.}(2022)\citenamefont {Yang},
  \citenamefont {Raymond},\ and\ \citenamefont {Uno}}]{yang2022efficient}%
  \BibitemOpen
  \bibfield  {author} {\bibinfo {author} {\bibfnamefont {Bo}~\bibnamefont
  {Yang}}, \bibinfo {author} {\bibfnamefont {Rudy}\ \bibnamefont {Raymond}}, \
  and\ \bibinfo {author} {\bibfnamefont {Shumpei}\ \bibnamefont {Uno}},\
  }\bibfield  {title} {\enquote {\bibinfo {title} {Efficient quantum
  readout-error mitigation for sparse measurement outcomes of near-term quantum
  devices},}\ }\href@noop {} {\bibfield  {journal} {\bibinfo  {journal}
  {Physical Review A}\ }\textbf {\bibinfo {volume} {106}},\ \bibinfo {pages}
  {012423} (\bibinfo {year} {2022})}\BibitemShut {NoStop}%
\bibitem [{\citenamefont {Zlokapa}\ and\ \citenamefont
  {Gheorghiu}(2020)}]{Zlo20}%
  \BibitemOpen
  \bibfield  {author} {\bibinfo {author} {\bibfnamefont {A.}~\bibnamefont
  {Zlokapa}}\ and\ \bibinfo {author} {\bibfnamefont {A.}~\bibnamefont
  {Gheorghiu}},\ }\href@noop {} {\enquote {\bibinfo {title} {A deep learning
  model for noise prediction on near-term quantum devices},}\ } (\bibinfo
  {year} {2020}),\ \Eprint {http://arxiv.org/abs/2005.10811} {arXiv:2005.10811}
  \BibitemShut {NoStop}%
\bibitem [{\citenamefont {Das}\ \emph {et~al.}(2021)\citenamefont {Das},
  \citenamefont {Tannu}, \citenamefont {Dangwal},\ and\ \citenamefont
  {Qureshi}}]{Das21}%
  \BibitemOpen
  \bibfield  {author} {\bibinfo {author} {\bibfnamefont {P.}~\bibnamefont
  {Das}}, \bibinfo {author} {\bibfnamefont {S.}~\bibnamefont {Tannu}}, \bibinfo
  {author} {\bibfnamefont {S.}~\bibnamefont {Dangwal}}, \ and\ \bibinfo
  {author} {\bibfnamefont {M.}~\bibnamefont {Qureshi}},\ }\bibfield  {title}
  {\enquote {\bibinfo {title} {Adapt: Mitigating idling errors in qubits via
  adaptive dynamical decoupling},}\ }in\ \href@noop {} {\emph {\bibinfo
  {booktitle} {MICRO-54: 54th Annual IEEE/ACM International Symposium on
  Microarchitecture}}}\ (\bibinfo {year} {2021})\ pp.\ \bibinfo {pages}
  {950--962}\BibitemShut {NoStop}%
\bibitem [{\citenamefont {Ravi}\ \emph {et~al.}(2022)\citenamefont {Ravi},
  \citenamefont {Smith}, \citenamefont {Gokhale}, \citenamefont {Mari},
  \citenamefont {Earnest}, \citenamefont {Javadi-Abhari},\ and\ \citenamefont
  {Chong}}]{Rav22}%
  \BibitemOpen
  \bibfield  {author} {\bibinfo {author} {\bibfnamefont {G.~K.}\ \bibnamefont
  {Ravi}}, \bibinfo {author} {\bibfnamefont {K.~N.}\ \bibnamefont {Smith}},
  \bibinfo {author} {\bibfnamefont {P.}~\bibnamefont {Gokhale}}, \bibinfo
  {author} {\bibfnamefont {A.}~\bibnamefont {Mari}}, \bibinfo {author}
  {\bibfnamefont {N.}~\bibnamefont {Earnest}}, \bibinfo {author} {\bibfnamefont
  {A.}~\bibnamefont {Javadi-Abhari}}, \ and\ \bibinfo {author} {\bibfnamefont
  {F.~T.}\ \bibnamefont {Chong}},\ }\bibfield  {title} {\enquote {\bibinfo
  {title} {{VAQEM: A Variational Approach to Quantum Error Mitigation}},}\ }in\
  \href {\doibase 10.1109/HPCA53966.2022.00029} {\emph {\bibinfo {booktitle}
  {2022 IEEE International Symposium on High-Performance Computer Architecture
  (HPCA)}}}\ (\bibinfo {year} {2022})\ pp.\ \bibinfo {pages}
  {288--303}\BibitemShut {NoStop}%
\bibitem [{\citenamefont {Quiroz}\ and\ \citenamefont {Lidar}(2013)}]{Qui13}%
  \BibitemOpen
  \bibfield  {author} {\bibinfo {author} {\bibfnamefont {G.}~\bibnamefont
  {Quiroz}}\ and\ \bibinfo {author} {\bibfnamefont {D.~A.}\ \bibnamefont
  {Lidar}},\ }\bibfield  {title} {\enquote {\bibinfo {title} {Optimized
  dynamical decoupling via genetic algorithms},}\ }\href@noop {} {\bibfield
  {journal} {\bibinfo  {journal} {Physical Review A}\ }\textbf {\bibinfo
  {volume} {88}},\ \bibinfo {pages} {052306} (\bibinfo {year}
  {2013})}\BibitemShut {NoStop}%
\bibitem [{\citenamefont {Wu}\ and\ \citenamefont {Lidar}(2002)}]{Wul02}%
  \BibitemOpen
  \bibfield  {author} {\bibinfo {author} {\bibfnamefont {L.}~\bibnamefont
  {Wu}}\ and\ \bibinfo {author} {\bibfnamefont {D.~A.}\ \bibnamefont {Lidar}},\
  }\bibfield  {title} {\enquote {\bibinfo {title} {Creating decoherence-free
  subspaces using strong and fast pulses},}\ }\href {\doibase
  10.1103/PhysRevLett.88.207902} {\bibfield  {journal} {\bibinfo  {journal}
  {Phys. Rev. Lett.}\ }\textbf {\bibinfo {volume} {88}},\ \bibinfo {pages}
  {207902} (\bibinfo {year} {2002})}\BibitemShut {NoStop}%
\bibitem [{\citenamefont {Lidar}(2014)}]{Lid14}%
  \BibitemOpen
  \bibfield  {author} {\bibinfo {author} {\bibfnamefont {D.~A.}\ \bibnamefont
  {Lidar}},\ }\enquote {\bibinfo {title} {Review of decoherence-free subspaces,
  noiseless subsystems, and dynamical decoupling},}\ in\ \href {\doibase
  https://doi.org/10.1002/9781118742631.ch11} {\emph {\bibinfo {booktitle}
  {Quantum Information and Computation for Chemistry}}}\ (\bibinfo  {publisher}
  {John Wiley \& Sons, Ltd},\ \bibinfo {year} {2014})\ Chap.~\bibinfo {chapter}
  {11}, pp.\ \bibinfo {pages} {295--354}\BibitemShut {NoStop}%
\bibitem [{Note1()}]{Note1}%
  \BibitemOpen
  \bibinfo {note} {For a decoupling group of size $\abs {G}$ and pulse
  sequences of length $L$, there are $\protect \binom {\abs {G}+L-2}{\abs
  {G}-1}$ equivalence classes of DD sequences with distinct $U(T)$ operators
  when neglecting contributions to the effective system-bath Hamiltonian of
  $O(\tau ^2)$ and higher. This result arises from counting the number of ways
  that the $L-1$ intervals where the system evolves under the effective
  Hamiltonian $g_j^\protect \dag H_{SB} g_j$ can be partitioned among the $\abs
  {G}$ possibilities for $g_j$.}\BibitemShut {Stop}%
\bibitem [{\citenamefont {Proctor}\ \emph
  {et~al.}(2022{\natexlab{a}})\citenamefont {Proctor}, \citenamefont
  {Rudinger}, \citenamefont {Young}, \citenamefont {Nielsen},\ and\
  \citenamefont {Blume-Kohout}}]{Pro22_2}%
  \BibitemOpen
  \bibfield  {author} {\bibinfo {author} {\bibfnamefont {T.}~\bibnamefont
  {Proctor}}, \bibinfo {author} {\bibfnamefont {K.}~\bibnamefont {Rudinger}},
  \bibinfo {author} {\bibfnamefont {K.}~\bibnamefont {Young}}, \bibinfo
  {author} {\bibfnamefont {E.}~\bibnamefont {Nielsen}}, \ and\ \bibinfo
  {author} {\bibfnamefont {R.}~\bibnamefont {Blume-Kohout}},\ }\bibfield
  {title} {\enquote {\bibinfo {title} {Measuring the capabilities of quantum
  computers},}\ }\href@noop {} {\bibfield  {journal} {\bibinfo  {journal}
  {Nature Physics}\ }\textbf {\bibinfo {volume} {18}},\ \bibinfo {pages}
  {75--79} (\bibinfo {year} {2022}{\natexlab{a}})}\BibitemShut {NoStop}%
\bibitem [{\citenamefont {Aaronson}\ and\ \citenamefont
  {Gottesman}(2004)}]{Aar04}%
  \BibitemOpen
  \bibfield  {author} {\bibinfo {author} {\bibfnamefont {S.}~\bibnamefont
  {Aaronson}}\ and\ \bibinfo {author} {\bibfnamefont {D.}~\bibnamefont
  {Gottesman}},\ }\bibfield  {title} {\enquote {\bibinfo {title} {Improved
  simulation of stabilizer circuits},}\ }\href {\doibase
  10.1103/PhysRevA.70.052328} {\bibfield  {journal} {\bibinfo  {journal} {Phys.
  Rev. A}\ }\textbf {\bibinfo {volume} {70}},\ \bibinfo {pages} {052328}
  (\bibinfo {year} {2004})}\BibitemShut {NoStop}%
\bibitem [{\citenamefont {Reeves}\ and\ \citenamefont
  {Rowe}(2002)}]{reeves2002genetic}%
  \BibitemOpen
  \bibfield  {author} {\bibinfo {author} {\bibfnamefont {C.}~\bibnamefont
  {Reeves}}\ and\ \bibinfo {author} {\bibfnamefont {J.~E.}\ \bibnamefont
  {Rowe}},\ }\href@noop {} {\emph {\bibinfo {title} {Genetic algorithms:
  principles and perspectives: a guide to GA theory}}},\ Vol.~\bibinfo {volume}
  {20}\ (\bibinfo  {publisher} {Springer Science \& Business Media},\ \bibinfo
  {year} {2002})\BibitemShut {NoStop}%
\bibitem [{\citenamefont {Mitchell}(1999)}]{Mit99}%
  \BibitemOpen
  \bibfield  {author} {\bibinfo {author} {\bibfnamefont {M.}~\bibnamefont
  {Mitchell}},\ }\href@noop {} {\emph {\bibinfo {title} {An Introduction to
  Genetic Algorithms}}}\ (\bibinfo  {publisher} {MIT Press},\ \bibinfo
  {address} {Cambridge},\ \bibinfo {year} {1999})\BibitemShut {NoStop}%
\bibitem [{Note2()}]{Note2}%
  \BibitemOpen
  \bibinfo {note} {$I_m$ and $I_p$ both act as the identity element, although
  $I_m$ is included for purposes of group structure}\BibitemShut {NoStop}%
\bibitem [{\citenamefont {Chamberland}\ \emph {et~al.}(2020)\citenamefont
  {Chamberland}, \citenamefont {Zhu}, \citenamefont {Yoder}, \citenamefont
  {Hertzberg},\ and\ \citenamefont {Cross}}]{Cha20}%
  \BibitemOpen
  \bibfield  {author} {\bibinfo {author} {\bibfnamefont {C.}~\bibnamefont
  {Chamberland}}, \bibinfo {author} {\bibfnamefont {G.}~\bibnamefont {Zhu}},
  \bibinfo {author} {\bibfnamefont {T.~J.}\ \bibnamefont {Yoder}}, \bibinfo
  {author} {\bibfnamefont {J.~B.}\ \bibnamefont {Hertzberg}}, \ and\ \bibinfo
  {author} {\bibfnamefont {A.~W.}\ \bibnamefont {Cross}},\ }\bibfield  {title}
  {\enquote {\bibinfo {title} {Topological and subsystem codes on low-degree
  graphs with flag qubits},}\ }\href {\doibase 10.1103/PhysRevX.10.011022}
  {\bibfield  {journal} {\bibinfo  {journal} {Phys. Rev. X}\ }\textbf {\bibinfo
  {volume} {10}},\ \bibinfo {pages} {011022} (\bibinfo {year}
  {2020})}\BibitemShut {NoStop}%
\bibitem [{Note3()}]{Note3}%
  \BibitemOpen
  \bibinfo {note} {To impose that such allocations of identity operations
  indeed lead to distinct staggerings, the DD sequence associated with the
  first color in the strategy has the pulse sequence placed uniformly between
  the operations defining the start and end of the idle period, while the
  second receives an asymmetric sequence with a pulse at the earliest time one
  can be applied in the idle period and the third receives an asymmetric
  sequence with a pulse at the latest time one can be applied in the idle
  period.}\BibitemShut {Stop}%
\bibitem [{\citenamefont {Polak}(2012)}]{Pol2012}%
  \BibitemOpen
  \bibfield  {author} {\bibinfo {author} {\bibfnamefont {E.}~\bibnamefont
  {Polak}},\ }\href@noop {} {\emph {\bibinfo {title} {Optimization: algorithms
  and consistent approximations}}},\ Vol.\ \bibinfo {volume} {124}\ (\bibinfo
  {publisher} {Springer Science \& Business Media},\ \bibinfo {year}
  {2012})\BibitemShut {NoStop}%
\bibitem [{\citenamefont {Egger}\ \emph {et~al.}(2023)\citenamefont {Egger},
  \citenamefont {Capecci}, \citenamefont {Pokharel}, \citenamefont
  {Barkoutsos}, \citenamefont {Fischer}, \citenamefont {Guidoni},\ and\
  \citenamefont {Tavernelli}}]{Egg23}%
  \BibitemOpen
  \bibfield  {author} {\bibinfo {author} {\bibfnamefont {D.~J.}\ \bibnamefont
  {Egger}}, \bibinfo {author} {\bibfnamefont {C.}~\bibnamefont {Capecci}},
  \bibinfo {author} {\bibfnamefont {B.}~\bibnamefont {Pokharel}}, \bibinfo
  {author} {\bibfnamefont {P.~K.}\ \bibnamefont {Barkoutsos}}, \bibinfo
  {author} {\bibfnamefont {L.~E.}\ \bibnamefont {Fischer}}, \bibinfo {author}
  {\bibfnamefont {L.}~\bibnamefont {Guidoni}}, \ and\ \bibinfo {author}
  {\bibfnamefont {I.}~\bibnamefont {Tavernelli}},\ }\bibfield  {title}
  {\enquote {\bibinfo {title} {Pulse variational quantum eigensolver on
  cross-resonance-based hardware},}\ }\href {\doibase
  10.1103/PhysRevResearch.5.033159} {\bibfield  {journal} {\bibinfo  {journal}
  {Phys. Rev. Res.}\ }\textbf {\bibinfo {volume} {5}},\ \bibinfo {pages}
  {033159} (\bibinfo {year} {2023})}\BibitemShut {NoStop}%
\bibitem [{\citenamefont {Rahman}\ \emph {et~al.}(2024)\citenamefont {Rahman},
  \citenamefont {Egger},\ and\ \citenamefont {Arenz}}]{Rah24}%
  \BibitemOpen
  \bibfield  {author} {\bibinfo {author} {\bibfnamefont {Arefur}\ \bibnamefont
  {Rahman}}, \bibinfo {author} {\bibfnamefont {Daniel~J.}\ \bibnamefont
  {Egger}}, \ and\ \bibinfo {author} {\bibfnamefont {Christian}\ \bibnamefont
  {Arenz}},\ }\href {https://arxiv.org/abs/2405.08689} {\enquote {\bibinfo
  {title} {Learning how to dynamically decouple},}\ } (\bibinfo {year}
  {2024}),\ \bibinfo {note} {2405.08689}\BibitemShut {NoStop}%
\bibitem [{\citenamefont {Bernstein}\ and\ \citenamefont
  {Vazirani}(1993)}]{Ber93}%
  \BibitemOpen
  \bibfield  {author} {\bibinfo {author} {\bibfnamefont {E.}~\bibnamefont
  {Bernstein}}\ and\ \bibinfo {author} {\bibfnamefont {U.}~\bibnamefont
  {Vazirani}},\ }\bibfield  {title} {\enquote {\bibinfo {title} {Quantum
  complexity theory},}\ }in\ \href@noop {} {\emph {\bibinfo {booktitle}
  {Proceedings of the twenty-fifth annual ACM symposium on Theory of
  computing}}}\ (\bibinfo {year} {1993})\ pp.\ \bibinfo {pages}
  {11--20}\BibitemShut {NoStop}%
\bibitem [{\citenamefont {Lubinski}\ \emph {et~al.}(2023)\citenamefont
  {Lubinski}, \citenamefont {Johri}, \citenamefont {Varosy}, \citenamefont
  {Coleman}, \citenamefont {Zhao}, \citenamefont {Necaise}, \citenamefont
  {Baldwin}, \citenamefont {Mayer},\ and\ \citenamefont
  {Proctor}}]{lubinski2023application}%
  \BibitemOpen
  \bibfield  {author} {\bibinfo {author} {\bibfnamefont {T.}~\bibnamefont
  {Lubinski}}, \bibinfo {author} {\bibfnamefont {S.}~\bibnamefont {Johri}},
  \bibinfo {author} {\bibfnamefont {P.}~\bibnamefont {Varosy}}, \bibinfo
  {author} {\bibfnamefont {J.}~\bibnamefont {Coleman}}, \bibinfo {author}
  {\bibfnamefont {L.}~\bibnamefont {Zhao}}, \bibinfo {author} {\bibfnamefont
  {J.}~\bibnamefont {Necaise}}, \bibinfo {author} {\bibfnamefont {C.~H.}\
  \bibnamefont {Baldwin}}, \bibinfo {author} {\bibfnamefont {K.}~\bibnamefont
  {Mayer}}, \ and\ \bibinfo {author} {\bibfnamefont {T.}~\bibnamefont
  {Proctor}},\ }\bibfield  {title} {\enquote {\bibinfo {title}
  {Application-oriented performance benchmarks for quantum computing},}\
  }\href@noop {} {\bibfield  {journal} {\bibinfo  {journal} {IEEE Transactions
  on Quantum Engineering}\ } (\bibinfo {year} {2023})}\BibitemShut {NoStop}%
\bibitem [{\citenamefont {Fallek}\ \emph {et~al.}(2016)\citenamefont {Fallek},
  \citenamefont {Herold}, \citenamefont {McMahon}, \citenamefont {Maller},
  \citenamefont {Brown},\ and\ \citenamefont {Amini}}]{fallek2016transport}%
  \BibitemOpen
  \bibfield  {author} {\bibinfo {author} {\bibfnamefont {S.}~\bibnamefont
  {Fallek}}, \bibinfo {author} {\bibfnamefont {C.}~\bibnamefont {Herold}},
  \bibinfo {author} {\bibfnamefont {B.}~\bibnamefont {McMahon}}, \bibinfo
  {author} {\bibfnamefont {K.}~\bibnamefont {Maller}}, \bibinfo {author}
  {\bibfnamefont {K.}~\bibnamefont {Brown}}, \ and\ \bibinfo {author}
  {\bibfnamefont {J.}~\bibnamefont {Amini}},\ }\bibfield  {title} {\enquote
  {\bibinfo {title} {Transport implementation of the {Bernstein}--{Vazirani}
  algorithm with ion qubits},}\ }\href@noop {} {\bibfield  {journal} {\bibinfo
  {journal} {New Journal of Physics}\ }\textbf {\bibinfo {volume} {18}},\
  \bibinfo {pages} {083030} (\bibinfo {year} {2016})}\BibitemShut {NoStop}%
\bibitem [{\citenamefont {Wright}\ \emph {et~al.}(2019)\citenamefont {Wright},
  \citenamefont {Beck}, \citenamefont {Debnath}, \citenamefont {Amini},
  \citenamefont {Nam}, \citenamefont {Grzesiak}, \citenamefont {Chen},
  \citenamefont {Pisenti}, \citenamefont {Chmielewski}, \citenamefont {Collins}
  \emph {et~al.}}]{wright2019benchmarking}%
  \BibitemOpen
  \bibfield  {author} {\bibinfo {author} {\bibfnamefont {K.}~\bibnamefont
  {Wright}}, \bibinfo {author} {\bibfnamefont {K.~M.}\ \bibnamefont {Beck}},
  \bibinfo {author} {\bibfnamefont {S.}~\bibnamefont {Debnath}}, \bibinfo
  {author} {\bibfnamefont {J.}~\bibnamefont {Amini}}, \bibinfo {author}
  {\bibfnamefont {Y.}~\bibnamefont {Nam}}, \bibinfo {author} {\bibfnamefont
  {N.}~\bibnamefont {Grzesiak}}, \bibinfo {author} {\bibfnamefont
  {J.}~\bibnamefont {Chen}}, \bibinfo {author} {\bibfnamefont {N.}~\bibnamefont
  {Pisenti}}, \bibinfo {author} {\bibfnamefont {M.}~\bibnamefont
  {Chmielewski}}, \bibinfo {author} {\bibfnamefont {C.}~\bibnamefont
  {Collins}},  \emph {et~al.},\ }\bibfield  {title} {\enquote {\bibinfo {title}
  {Benchmarking an 11-qubit quantum computer},}\ }\href@noop {} {\bibfield
  {journal} {\bibinfo  {journal} {Nature Communications}\ }\textbf {\bibinfo
  {volume} {10}},\ \bibinfo {pages} {5464} (\bibinfo {year}
  {2019})}\BibitemShut {NoStop}%
\bibitem [{\citenamefont {{The Qiskit Research developers and
  contributors}}(2023)}]{qiskit_research}%
  \BibitemOpen
  \bibfield  {author} {\bibinfo {author} {\bibnamefont {{The Qiskit Research
  developers and contributors}}},\ }\href {\doibase 10.5281/zenodo.7776174}
  {\enquote {\bibinfo {title} {{Qiskit Research}},}\ } (\bibinfo {year}
  {2023})\BibitemShut {NoStop}%
\bibitem [{\citenamefont {Mundada}\ \emph {et~al.}(2023)\citenamefont
  {Mundada}, \citenamefont {Barbosa}, \citenamefont {Maity}, \citenamefont
  {Wang}, \citenamefont {Merkh}, \citenamefont {Stace}, \citenamefont
  {Nielson}, \citenamefont {Carvalho}, \citenamefont {Hush}, \citenamefont
  {Biercuk} \emph {et~al.}}]{Mun23}%
  \BibitemOpen
  \bibfield  {author} {\bibinfo {author} {\bibfnamefont {P.~S.}\ \bibnamefont
  {Mundada}}, \bibinfo {author} {\bibfnamefont {A.}~\bibnamefont {Barbosa}},
  \bibinfo {author} {\bibfnamefont {S.}~\bibnamefont {Maity}}, \bibinfo
  {author} {\bibfnamefont {Y.}~\bibnamefont {Wang}}, \bibinfo {author}
  {\bibfnamefont {T.}~\bibnamefont {Merkh}}, \bibinfo {author} {\bibfnamefont
  {T.}~\bibnamefont {Stace}}, \bibinfo {author} {\bibfnamefont
  {F.}~\bibnamefont {Nielson}}, \bibinfo {author} {\bibfnamefont {A.~R.}\
  \bibnamefont {Carvalho}}, \bibinfo {author} {\bibfnamefont {M.}~\bibnamefont
  {Hush}}, \bibinfo {author} {\bibfnamefont {M.~J.}\ \bibnamefont {Biercuk}},
  \emph {et~al.},\ }\bibfield  {title} {\enquote {\bibinfo {title}
  {Experimental benchmarking of an automated deterministic error-suppression
  workflow for quantum algorithms},}\ }\href@noop {} {\bibfield  {journal}
  {\bibinfo  {journal} {Physical Review Applied}\ }\textbf {\bibinfo {volume}
  {20}},\ \bibinfo {pages} {024034} (\bibinfo {year} {2023})}\BibitemShut
  {NoStop}%
\bibitem [{\citenamefont {Lidar}\ and\ \citenamefont
  {Brun}(2013)}]{lidar2013quantum}%
  \BibitemOpen
  \bibfield  {author} {\bibinfo {author} {\bibfnamefont {Daniel~A}\
  \bibnamefont {Lidar}}\ and\ \bibinfo {author} {\bibfnamefont {Todd~A}\
  \bibnamefont {Brun}},\ }\href@noop {} {\emph {\bibinfo {title} {Quantum error
  correction}}}\ (\bibinfo  {publisher} {Cambridge university press},\ \bibinfo
  {year} {2013})\BibitemShut {NoStop}%
\bibitem [{\citenamefont {Greenberger}\ \emph {et~al.}(1989)\citenamefont
  {Greenberger}, \citenamefont {Horne},\ and\ \citenamefont
  {Zeilinger}}]{Gre89}%
  \BibitemOpen
  \bibfield  {author} {\bibinfo {author} {\bibfnamefont {Daniel~M.}\
  \bibnamefont {Greenberger}}, \bibinfo {author} {\bibfnamefont {Michael~A.}\
  \bibnamefont {Horne}}, \ and\ \bibinfo {author} {\bibfnamefont {Anton}\
  \bibnamefont {Zeilinger}},\ }\enquote {\bibinfo {title} {Going beyond bell's
  theorem},}\ in\ \href {\doibase 10.1007/978-94-017-0849-4_10} {\emph
  {\bibinfo {booktitle} {Bell's Theorem, Quantum Theory and Conceptions of the
  Universe}}},\ \bibinfo {editor} {edited by\ \bibinfo {editor} {\bibfnamefont
  {Menas}\ \bibnamefont {Kafatos}}}\ (\bibinfo  {publisher} {Springer
  Netherlands},\ \bibinfo {address} {Dordrecht},\ \bibinfo {year} {1989})\ pp.\
  \bibinfo {pages} {69--72}\BibitemShut {NoStop}%
\bibitem [{\citenamefont {Hillery}\ \emph {et~al.}(1999)\citenamefont
  {Hillery}, \citenamefont {Bu\ifmmode~\check{z}\else \v{z}\fi{}ek},\ and\
  \citenamefont {Berthiaume}}]{Hil99}%
  \BibitemOpen
  \bibfield  {author} {\bibinfo {author} {\bibfnamefont {Mark}\ \bibnamefont
  {Hillery}}, \bibinfo {author} {\bibfnamefont {Vladim\'{\i}r}\ \bibnamefont
  {Bu\ifmmode~\check{z}\else \v{z}\fi{}ek}}, \ and\ \bibinfo {author}
  {\bibfnamefont {Andr\'e}\ \bibnamefont {Berthiaume}},\ }\bibfield  {title}
  {\enquote {\bibinfo {title} {Quantum secret sharing},}\ }\href {\doibase
  10.1103/PhysRevA.59.1829} {\bibfield  {journal} {\bibinfo  {journal} {Phys.
  Rev. A}\ }\textbf {\bibinfo {volume} {59}},\ \bibinfo {pages} {1829--1834}
  (\bibinfo {year} {1999})}\BibitemShut {NoStop}%
\bibitem [{\citenamefont {Omkar}\ \emph {et~al.}(2022)\citenamefont {Omkar},
  \citenamefont {Lee}, \citenamefont {Teo}, \citenamefont {Lee},\ and\
  \citenamefont {Jeong}}]{Omk22}%
  \BibitemOpen
  \bibfield  {author} {\bibinfo {author} {\bibfnamefont {Srikrishna}\
  \bibnamefont {Omkar}}, \bibinfo {author} {\bibfnamefont {Seok-Hyung}\
  \bibnamefont {Lee}}, \bibinfo {author} {\bibfnamefont {Yong~Siah}\
  \bibnamefont {Teo}}, \bibinfo {author} {\bibfnamefont {Seung-Woo}\
  \bibnamefont {Lee}}, \ and\ \bibinfo {author} {\bibfnamefont {Hyunseok}\
  \bibnamefont {Jeong}},\ }\bibfield  {title} {\enquote {\bibinfo {title}
  {All-photonic architecture for scalable quantum computing with
  greenberger-horne-zeilinger states},}\ }\href {\doibase
  10.1103/PRXQuantum.3.030309} {\bibfield  {journal} {\bibinfo  {journal} {PRX
  Quantum}\ }\textbf {\bibinfo {volume} {3}},\ \bibinfo {pages} {030309}
  (\bibinfo {year} {2022})}\BibitemShut {NoStop}%
\bibitem [{\citenamefont {Pankovich}\ \emph {et~al.}(2024)\citenamefont
  {Pankovich}, \citenamefont {Kan}, \citenamefont {Wan}, \citenamefont
  {Ostmann}, \citenamefont {Neville}, \citenamefont {Omkar}, \citenamefont
  {Sohbi},\ and\ \citenamefont {Br\'adler}}]{Pan24}%
  \BibitemOpen
  \bibfield  {author} {\bibinfo {author} {\bibfnamefont {Brendan}\ \bibnamefont
  {Pankovich}}, \bibinfo {author} {\bibfnamefont {Angus}\ \bibnamefont {Kan}},
  \bibinfo {author} {\bibfnamefont {Kwok~Ho}\ \bibnamefont {Wan}}, \bibinfo
  {author} {\bibfnamefont {Maike}\ \bibnamefont {Ostmann}}, \bibinfo {author}
  {\bibfnamefont {Alex}\ \bibnamefont {Neville}}, \bibinfo {author}
  {\bibfnamefont {Srikrishna}\ \bibnamefont {Omkar}}, \bibinfo {author}
  {\bibfnamefont {Adel}\ \bibnamefont {Sohbi}}, \ and\ \bibinfo {author}
  {\bibfnamefont {Kamil}\ \bibnamefont {Br\'adler}},\ }\bibfield  {title}
  {\enquote {\bibinfo {title} {High-photon-loss threshold quantum computing
  using ghz-state measurements},}\ }\href {\doibase
  10.1103/PhysRevLett.133.050604} {\bibfield  {journal} {\bibinfo  {journal}
  {Phys. Rev. Lett.}\ }\textbf {\bibinfo {volume} {133}},\ \bibinfo {pages}
  {050604} (\bibinfo {year} {2024})}\BibitemShut {NoStop}%
\bibitem [{\citenamefont {Chen}\ and\ \citenamefont {Lo}(2008)}]{Che08}%
  \BibitemOpen
  \bibfield  {author} {\bibinfo {author} {\bibfnamefont {Kai}\ \bibnamefont
  {Chen}}\ and\ \bibinfo {author} {\bibfnamefont {Hoi-Kwong}\ \bibnamefont
  {Lo}},\ }\href {https://arxiv.org/abs/quant-ph/0404133} {\enquote {\bibinfo
  {title} {Multi-partite quantum cryptographic protocols with noisy ghz
  states},}\ } (\bibinfo {year} {2008}),\ \Eprint
  {http://arxiv.org/abs/quant-ph/0404133} {arXiv:quant-ph/0404133 [quant-ph]}
  \BibitemShut {NoStop}%
\bibitem [{\citenamefont {Song}\ \emph {et~al.}(2019)\citenamefont {Song},
  \citenamefont {Xu}, \citenamefont {Li}, \citenamefont {Zhang}, \citenamefont
  {Zhang}, \citenamefont {Liu}, \citenamefont {Guo}, \citenamefont {Wang},
  \citenamefont {Ren}, \citenamefont {Hao}, \citenamefont {Feng}, \citenamefont
  {Fan}, \citenamefont {Zheng}, \citenamefont {Wang}, \citenamefont {Wang},\
  and\ \citenamefont {Zhu}}]{Son19}%
  \BibitemOpen
  \bibfield  {author} {\bibinfo {author} {\bibfnamefont {Chao}\ \bibnamefont
  {Song}}, \bibinfo {author} {\bibfnamefont {Kai}\ \bibnamefont {Xu}}, \bibinfo
  {author} {\bibfnamefont {Hekang}\ \bibnamefont {Li}}, \bibinfo {author}
  {\bibfnamefont {Yu-Ran}\ \bibnamefont {Zhang}}, \bibinfo {author}
  {\bibfnamefont {Xu}~\bibnamefont {Zhang}}, \bibinfo {author} {\bibfnamefont
  {Wuxin}\ \bibnamefont {Liu}}, \bibinfo {author} {\bibfnamefont {Qiujiang}\
  \bibnamefont {Guo}}, \bibinfo {author} {\bibfnamefont {Zhen}\ \bibnamefont
  {Wang}}, \bibinfo {author} {\bibfnamefont {Wenhui}\ \bibnamefont {Ren}},
  \bibinfo {author} {\bibfnamefont {Jie}\ \bibnamefont {Hao}}, \bibinfo
  {author} {\bibfnamefont {Hui}\ \bibnamefont {Feng}}, \bibinfo {author}
  {\bibfnamefont {Heng}\ \bibnamefont {Fan}}, \bibinfo {author} {\bibfnamefont
  {Dongning}\ \bibnamefont {Zheng}}, \bibinfo {author} {\bibfnamefont {Da-Wei}\
  \bibnamefont {Wang}}, \bibinfo {author} {\bibfnamefont {H.}~\bibnamefont
  {Wang}}, \ and\ \bibinfo {author} {\bibfnamefont {Shi-Yao}\ \bibnamefont
  {Zhu}},\ }\bibfield  {title} {\enquote {\bibinfo {title} {Generation of
  multicomponent atomic schrödinger cat states of up to 20 qubits},}\ }\href
  {\doibase 10.1126/science.aay0600} {\bibfield  {journal} {\bibinfo  {journal}
  {Science}\ }\textbf {\bibinfo {volume} {365}},\ \bibinfo {pages} {574--577}
  (\bibinfo {year} {2019})},\ \Eprint
  {http://arxiv.org/abs/https://www.science.org/doi/pdf/10.1126/science.aay0600}
  {https://www.science.org/doi/pdf/10.1126/science.aay0600} \BibitemShut
  {NoStop}%
\bibitem [{\citenamefont {Chen}\ \emph {et~al.}(2023)\citenamefont {Chen},
  \citenamefont {Zhu}, \citenamefont {Verresen}, \citenamefont {Seif},
  \citenamefont {Bäumer}, \citenamefont {Layden}, \citenamefont
  {Tantivasadakarn}, \citenamefont {Zhu}, \citenamefont {Sheldon},
  \citenamefont {Vishwanath}, \citenamefont {Trebst},\ and\ \citenamefont
  {Kandala}}]{Che23}%
  \BibitemOpen
  \bibfield  {author} {\bibinfo {author} {\bibfnamefont {Edward~H.}\
  \bibnamefont {Chen}}, \bibinfo {author} {\bibfnamefont {Guo-Yi}\ \bibnamefont
  {Zhu}}, \bibinfo {author} {\bibfnamefont {Ruben}\ \bibnamefont {Verresen}},
  \bibinfo {author} {\bibfnamefont {Alireza}\ \bibnamefont {Seif}}, \bibinfo
  {author} {\bibfnamefont {Elisa}\ \bibnamefont {Bäumer}}, \bibinfo {author}
  {\bibfnamefont {David}\ \bibnamefont {Layden}}, \bibinfo {author}
  {\bibfnamefont {Nathanan}\ \bibnamefont {Tantivasadakarn}}, \bibinfo {author}
  {\bibfnamefont {Guanyu}\ \bibnamefont {Zhu}}, \bibinfo {author}
  {\bibfnamefont {Sarah}\ \bibnamefont {Sheldon}}, \bibinfo {author}
  {\bibfnamefont {Ashvin}\ \bibnamefont {Vishwanath}}, \bibinfo {author}
  {\bibfnamefont {Simon}\ \bibnamefont {Trebst}}, \ and\ \bibinfo {author}
  {\bibfnamefont {Abhinav}\ \bibnamefont {Kandala}},\ }\href@noop {} {\enquote
  {\bibinfo {title} {Realizing the nishimori transition across the error
  threshold for constant-depth quantum circuits},}\ } (\bibinfo {year}
  {2023}),\ \Eprint {http://arxiv.org/abs/2309.02863} {arXiv:2309.02863
  [quant-ph]} \BibitemShut {NoStop}%
\bibitem [{\citenamefont {Magesan}\ \emph {et~al.}(2011)\citenamefont
  {Magesan}, \citenamefont {Gambetta},\ and\ \citenamefont
  {Emerson}}]{Mag2011}%
  \BibitemOpen
  \bibfield  {author} {\bibinfo {author} {\bibfnamefont {E.}~\bibnamefont
  {Magesan}}, \bibinfo {author} {\bibfnamefont {J.~M.}\ \bibnamefont
  {Gambetta}}, \ and\ \bibinfo {author} {\bibfnamefont {J.}~\bibnamefont
  {Emerson}},\ }\bibfield  {title} {\enquote {\bibinfo {title} {Scalable and
  robust randomized benchmarking of quantum processes},}\ }\href@noop {}
  {\bibfield  {journal} {\bibinfo  {journal} {Physical Review Letters}\
  }\textbf {\bibinfo {volume} {106}},\ \bibinfo {pages} {180504} (\bibinfo
  {year} {2011})}\BibitemShut {NoStop}%
\bibitem [{\citenamefont {Magesan}\ \emph {et~al.}(2012)\citenamefont
  {Magesan}, \citenamefont {Gambetta},\ and\ \citenamefont
  {Emerson}}]{Mag2012}%
  \BibitemOpen
  \bibfield  {author} {\bibinfo {author} {\bibfnamefont {E.}~\bibnamefont
  {Magesan}}, \bibinfo {author} {\bibfnamefont {J.~M.}\ \bibnamefont
  {Gambetta}}, \ and\ \bibinfo {author} {\bibfnamefont {J.}~\bibnamefont
  {Emerson}},\ }\bibfield  {title} {\enquote {\bibinfo {title} {Characterizing
  quantum gates via randomized benchmarking},}\ }\href@noop {} {\bibfield
  {journal} {\bibinfo  {journal} {Physical Review A}\ }\textbf {\bibinfo
  {volume} {85}},\ \bibinfo {pages} {042311} (\bibinfo {year}
  {2012})}\BibitemShut {NoStop}%
\bibitem [{\citenamefont {Proctor}\ \emph
  {et~al.}(2022{\natexlab{b}})\citenamefont {Proctor}, \citenamefont {Seritan},
  \citenamefont {Rudinger}, \citenamefont {Nielsen}, \citenamefont
  {Blume-Kohout},\ and\ \citenamefont {Young}}]{Pro22}%
  \BibitemOpen
  \bibfield  {author} {\bibinfo {author} {\bibfnamefont {T.}~\bibnamefont
  {Proctor}}, \bibinfo {author} {\bibfnamefont {S.}~\bibnamefont {Seritan}},
  \bibinfo {author} {\bibfnamefont {K.}~\bibnamefont {Rudinger}}, \bibinfo
  {author} {\bibfnamefont {E.}~\bibnamefont {Nielsen}}, \bibinfo {author}
  {\bibfnamefont {R.}~\bibnamefont {Blume-Kohout}}, \ and\ \bibinfo {author}
  {\bibfnamefont {K.}~\bibnamefont {Young}},\ }\bibfield  {title} {\enquote
  {\bibinfo {title} {Scalable randomized benchmarking of quantum computers
  using mirror circuits},}\ }\href@noop {} {\bibfield  {journal} {\bibinfo
  {journal} {Physical Review Letters}\ }\textbf {\bibinfo {volume} {129}},\
  \bibinfo {pages} {150502} (\bibinfo {year} {2022}{\natexlab{b}})}\BibitemShut
  {NoStop}%
\bibitem [{\citenamefont {Hines}\ \emph {et~al.}(2023)\citenamefont {Hines},
  \citenamefont {Lu}, \citenamefont {Naik}, \citenamefont {Hashim},
  \citenamefont {Ville}, \citenamefont {Mitchell}, \citenamefont {Kriekebaum},
  \citenamefont {Santiago}, \citenamefont {Seritan}, \citenamefont {Nielsen},
  \citenamefont {Blume-Kohout}, \citenamefont {Young}, \citenamefont {Siddiqi},
  \citenamefont {Whaley},\ and\ \citenamefont {Proctor}}]{Hin23}%
  \BibitemOpen
  \bibfield  {author} {\bibinfo {author} {\bibfnamefont {J.}~\bibnamefont
  {Hines}}, \bibinfo {author} {\bibfnamefont {M.}~\bibnamefont {Lu}}, \bibinfo
  {author} {\bibfnamefont {R.~K.}\ \bibnamefont {Naik}}, \bibinfo {author}
  {\bibfnamefont {A.}~\bibnamefont {Hashim}}, \bibinfo {author} {\bibfnamefont
  {J.}~\bibnamefont {Ville}}, \bibinfo {author} {\bibfnamefont
  {B.}~\bibnamefont {Mitchell}}, \bibinfo {author} {\bibfnamefont {J.~M.}\
  \bibnamefont {Kriekebaum}}, \bibinfo {author} {\bibfnamefont {D.~I.}\
  \bibnamefont {Santiago}}, \bibinfo {author} {\bibfnamefont {S.}~\bibnamefont
  {Seritan}}, \bibinfo {author} {\bibfnamefont {E.}~\bibnamefont {Nielsen}},
  \bibinfo {author} {\bibfnamefont {R.}~\bibnamefont {Blume-Kohout}}, \bibinfo
  {author} {\bibfnamefont {K.}~\bibnamefont {Young}}, \bibinfo {author}
  {\bibfnamefont {I.}~\bibnamefont {Siddiqi}}, \bibinfo {author} {\bibfnamefont
  {B.}~\bibnamefont {Whaley}}, \ and\ \bibinfo {author} {\bibfnamefont
  {T.}~\bibnamefont {Proctor}},\ }\bibfield  {title} {\enquote {\bibinfo
  {title} {Demonstrating scalable randomized benchmarking of universal gate
  sets},}\ }\href {\doibase 10.1103/PhysRevX.13.041030} {\bibfield  {journal}
  {\bibinfo  {journal} {Phys. Rev. X}\ }\textbf {\bibinfo {volume} {13}},\
  \bibinfo {pages} {041030} (\bibinfo {year} {2023})}\BibitemShut {NoStop}%
\bibitem [{\citenamefont {McKay}\ \emph {et~al.}(2023)\citenamefont {McKay},
  \citenamefont {Hincks}, \citenamefont {Pritchett}, \citenamefont {Carroll},
  \citenamefont {Govia},\ and\ \citenamefont {Merkel}}]{mckay2023}%
  \BibitemOpen
  \bibfield  {author} {\bibinfo {author} {\bibfnamefont {David~C}\ \bibnamefont
  {McKay}}, \bibinfo {author} {\bibfnamefont {Ian}\ \bibnamefont {Hincks}},
  \bibinfo {author} {\bibfnamefont {Emily~J}\ \bibnamefont {Pritchett}},
  \bibinfo {author} {\bibfnamefont {Malcolm}\ \bibnamefont {Carroll}}, \bibinfo
  {author} {\bibfnamefont {Luke~CG}\ \bibnamefont {Govia}}, \ and\ \bibinfo
  {author} {\bibfnamefont {Seth~T}\ \bibnamefont {Merkel}},\ }\bibfield
  {title} {\enquote {\bibinfo {title} {Benchmarking quantum processor
  performance at scale},}\ }\href@noop {} {\bibfield  {journal} {\bibinfo
  {journal} {arXiv preprint arXiv:2311.05933}\ } (\bibinfo {year}
  {2023})}\BibitemShut {NoStop}%
\bibitem [{\citenamefont {Amico}\ \emph {et~al.}(2023)\citenamefont {Amico},
  \citenamefont {Zhang}, \citenamefont {Jurcevic}, \citenamefont {Bishop},
  \citenamefont {Nation}, \citenamefont {Wack},\ and\ \citenamefont
  {McKay}}]{Ami23}%
  \BibitemOpen
  \bibfield  {author} {\bibinfo {author} {\bibfnamefont {Mirko}\ \bibnamefont
  {Amico}}, \bibinfo {author} {\bibfnamefont {Helena}\ \bibnamefont {Zhang}},
  \bibinfo {author} {\bibfnamefont {Petar}\ \bibnamefont {Jurcevic}}, \bibinfo
  {author} {\bibfnamefont {Lev~S.}\ \bibnamefont {Bishop}}, \bibinfo {author}
  {\bibfnamefont {Paul}\ \bibnamefont {Nation}}, \bibinfo {author}
  {\bibfnamefont {Andrew}\ \bibnamefont {Wack}}, \ and\ \bibinfo {author}
  {\bibfnamefont {David~C.}\ \bibnamefont {McKay}},\ }\bibfield  {title}
  {\enquote {\bibinfo {title} {Defining best practices for quantum
  benchmarks},}\ }in\ \href {\doibase 10.1109/QCE57702.2023.00084} {\emph
  {\bibinfo {booktitle} {2023 IEEE International Conference on Quantum
  Computing and Engineering (QCE)}}},\ Vol.~\bibinfo {volume} {01}\ (\bibinfo
  {year} {2023})\ pp.\ \bibinfo {pages} {692--702}\BibitemShut {NoStop}%
\bibitem [{\citenamefont {Malekakhlagh}\ \emph {et~al.}(2020)\citenamefont
  {Malekakhlagh}, \citenamefont {Magesan},\ and\ \citenamefont
  {McKay}}]{malekakhlagh2020first}%
  \BibitemOpen
  \bibfield  {author} {\bibinfo {author} {\bibfnamefont {Moein}\ \bibnamefont
  {Malekakhlagh}}, \bibinfo {author} {\bibfnamefont {Easwar}\ \bibnamefont
  {Magesan}}, \ and\ \bibinfo {author} {\bibfnamefont {David~C}\ \bibnamefont
  {McKay}},\ }\bibfield  {title} {\enquote {\bibinfo {title} {First-principles
  analysis of cross-resonance gate operation},}\ }\href@noop {} {\bibfield
  {journal} {\bibinfo  {journal} {Physical Review A}\ }\textbf {\bibinfo
  {volume} {102}},\ \bibinfo {pages} {042605} (\bibinfo {year}
  {2020})}\BibitemShut {NoStop}%
\bibitem [{\citenamefont {Santos}\ and\ \citenamefont {Viola}(2006)}]{Lea06}%
  \BibitemOpen
  \bibfield  {author} {\bibinfo {author} {\bibfnamefont {L.~F.}\ \bibnamefont
  {Santos}}\ and\ \bibinfo {author} {\bibfnamefont {L.}~\bibnamefont {Viola}},\
  }\bibfield  {title} {\enquote {\bibinfo {title} {Enhanced convergence and
  robust performance of randomized dynamical decoupling},}\ }\href@noop {}
  {\bibfield  {journal} {\bibinfo  {journal} {Phys. Rev. Lett.}\ }\textbf
  {\bibinfo {volume} {97}},\ \bibinfo {pages} {150501} (\bibinfo {year}
  {2006})}\BibitemShut {NoStop}%
\bibitem [{\citenamefont {Grover}(1997)}]{Gro97}%
  \BibitemOpen
  \bibfield  {author} {\bibinfo {author} {\bibfnamefont {L.~K.}\ \bibnamefont
  {Grover}},\ }\bibfield  {title} {\enquote {\bibinfo {title} {Quantum
  mechanics helps in searching for a needle in a haystack},}\ }\href {\doibase
  10.1103/PhysRevLett.79.325} {\bibfield  {journal} {\bibinfo  {journal} {Phys.
  Rev. Lett.}\ }\textbf {\bibinfo {volume} {79}},\ \bibinfo {pages} {325--328}
  (\bibinfo {year} {1997})}\BibitemShut {NoStop}%
\bibitem [{\citenamefont {{Qiskit Contributors}}(2023)}]{Qiskit}%
  \BibitemOpen
  \bibfield  {author} {\bibinfo {author} {\bibnamefont {{Qiskit
  Contributors}}},\ }\href {\doibase 10.5281/zenodo.2573505} {\enquote
  {\bibinfo {title} {{Qiskit: An Open-source Framework for Quantum
  Computing}},}\ } (\bibinfo {year} {2023})\BibitemShut {NoStop}%
\bibitem [{\citenamefont {Kanazawa}\ \emph {et~al.}(2023)\citenamefont
  {Kanazawa}, \citenamefont {Egger}, \citenamefont {Ben-Haim}, \citenamefont
  {Zhang}, \citenamefont {Shanks}, \citenamefont {Aleksandrowicz},\ and\
  \citenamefont {Wood}}]{Qiskit_Experiments}%
  \BibitemOpen
  \bibfield  {author} {\bibinfo {author} {\bibfnamefont {N.}~\bibnamefont
  {Kanazawa}}, \bibinfo {author} {\bibfnamefont {D.~J.}\ \bibnamefont {Egger}},
  \bibinfo {author} {\bibfnamefont {Y.}~\bibnamefont {Ben-Haim}}, \bibinfo
  {author} {\bibfnamefont {H.}~\bibnamefont {Zhang}}, \bibinfo {author}
  {\bibfnamefont {W.~E.}\ \bibnamefont {Shanks}}, \bibinfo {author}
  {\bibfnamefont {G.}~\bibnamefont {Aleksandrowicz}}, \ and\ \bibinfo {author}
  {\bibfnamefont {C.~J.}\ \bibnamefont {Wood}},\ }\bibfield  {title} {\enquote
  {\bibinfo {title} {{Qiskit Experiments: A Python package to characterize and
  calibrate quantum computers}},}\ }\href {\doibase 10.21105/joss.05329}
  {\bibfield  {journal} {\bibinfo  {journal} {Journal of Open Source Software}\
  }\textbf {\bibinfo {volume} {8}},\ \bibinfo {pages} {5329} (\bibinfo {year}
  {2023})}\BibitemShut {NoStop}%
\bibitem [{\citenamefont {Johnson}(2022)}]{qiskit_runtime}%
  \BibitemOpen
  \bibfield  {author} {\bibinfo {author} {\bibfnamefont {B.}~\bibnamefont
  {Johnson}},\ }\bibfield  {title} {\enquote {\bibinfo {title} {Qiskit runtime,
  a quantum-classical execution platform for cloud-accessible quantum
  computers},}\ }in\ \href@noop {} {\emph {\bibinfo {booktitle} {APS March
  Meeting Abstracts}}},\ Vol.\ \bibinfo {volume} {2022}\ (\bibinfo {year}
  {2022})\ pp.\ \bibinfo {pages} {T28--002}\BibitemShut {NoStop}%
\end{thebibliography}%
\appendix
\section{Genetic algorithm \label{ss:reproduction_mutation}}
We begin with an initial population of DD strategies with size $K$. The initial population is constructed by ensuring that for each qubit subset, each pulse appears at each site the same number of times. For example, one initial population of size $K = 16$ of $L = 8$ DD sequences with the decoupling group $G = \{I_p, I_m, X_p, X_m, Y_p, Y_m, Z_p, Z_m\}$ applied to a 3-colored qubit graph consists of following set of sequences, each of which is applied to all qubit colors:

\begin{itemize}
\item $I_pI_mX_pX_mY_pY_mZ_pZ_m$
\item $I_pX_pY_pZ_pI_mX_mY_mZ_m$
\item $I_mX_pX_mY_pY_mZ_pZ_mI_p$
\item $X_pY_pZ_pI_mX_mY_mZ_mI_p$
\item $X_pX_mY_pY_mZ_pZ_mI_pI_m$
\item $Y_pZ_pI_mX_mY_mZ_mI_pX_p$
\item $X_mY_pY_mZ_pZ_mI_pI_mX_p$
\item $Z_pI_mX_mY_mZ_mI_pX_pY_p$
\item $Y_pY_mZ_pZ_mI_pI_mX_pX_m$
\item $I_mX_mY_mZ_mI_pX_pY_pZ_p$
\item $Y_mZ_pZ_mI_pI_mX_pX_mY_p$
\item $X_mY_mZ_mI_pX_pY_pZ_pI_m$
\item $Z_pZ_mI_pI_mX_pX_mY_pY_m$
\item $Y_mZ_mI_pX_pY_pZ_pI_mX_m$
\item $Z_mI_pI_mX_pX_mY_pY_mZ_p$
\item $Z_mI_pX_pY_pZ_pI_mX_mY_m$
\end{itemize}
At any given index of the pulse sequence, each element $g_j\in G$ is present at the same nonzero frequency. As a result, there is no artificial initial preference of the GA population for any individual pulse to take on any specific $g_j$ over other elements of $G$. 

\begin{figure}
    \centering
    \includegraphics[width = 9cm]{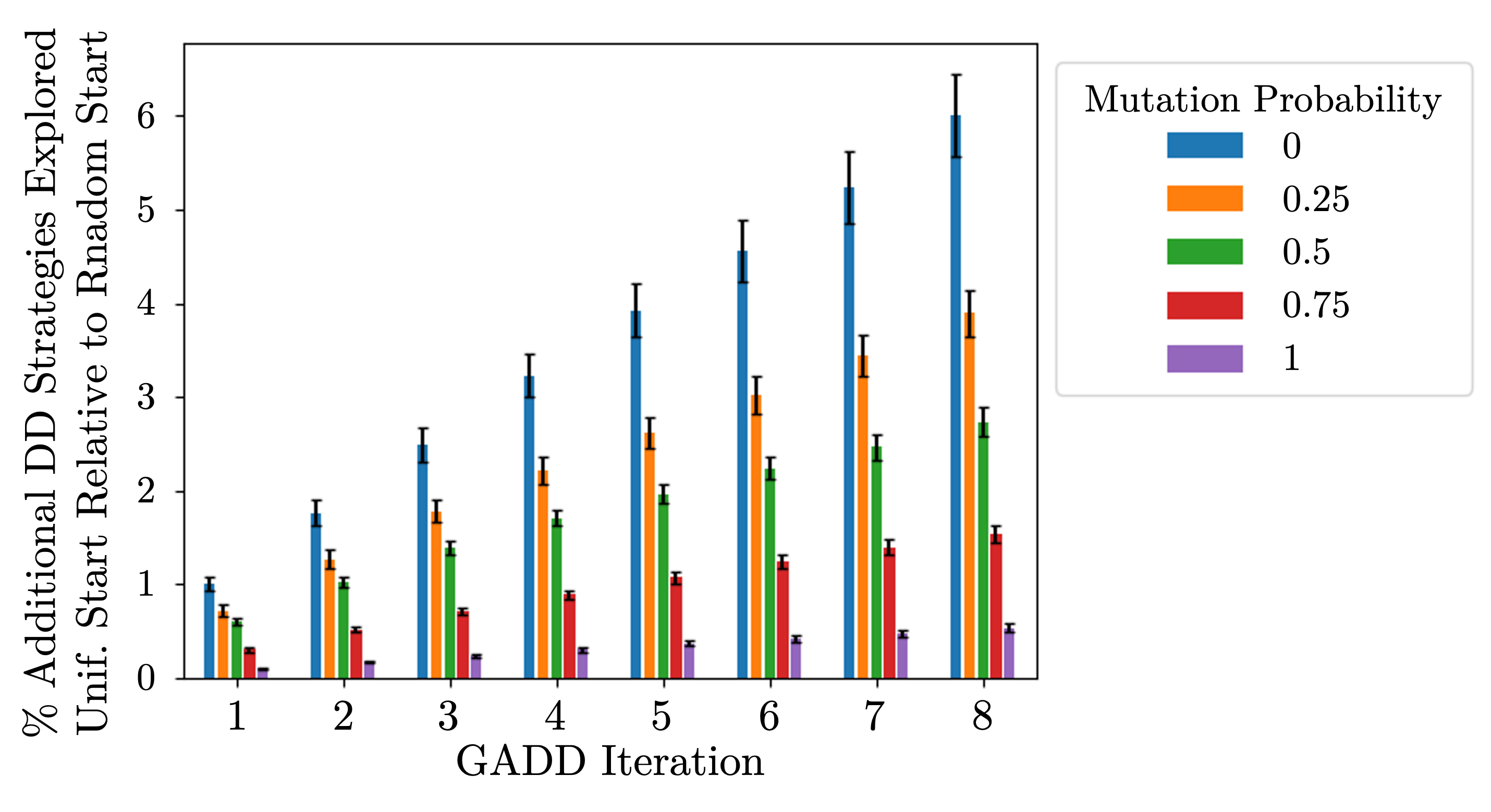}
    \caption{The GADD algorithm with $C = 1$ was used to explore the space of DD strategies on $G = \{I_p, I_m, X_p, X_m, Y_p, Y_m, Z_p, Z_m\}$ with utility functions assigned uniformly at random to all strategies in the case of the example uniform starting population and $25$ different randomly selected populations of $L = 8$ strategies. At each iteration, all pairs of DD strategies underwent reproduction to characterize the number of strategies that can theoretically be explored at that iteration number. Then, to maintain computational tractability, a random sample of 100 sequences was selected to be the new population. For each mutation probability, the average total number of DD sequences that can be explored in the random initial population case is compared relative to the average total number of DD sequences that can be explored in the uniform case. We find that at all GADD iteration numbers and mutation probabilities, starting the simulation from the example uniform starting population leads us to explore a larger amount of the DD strategy space compared to starting from a typical random population.}
    \label{fig:unif_start}
\end{figure}

We evaluate this method of constructing a uniform initial population by simulating the exploration of the genetic algorithm in the space of length 8 DD sequences when the utility associated with each sequence is uniformly selected at random on $[0, 1]$ to characterize how the GA explores sequences when it is agnostic of a utility function landscape. Since the DD sequence associated with any color evolves independently of other colors, this simulation is indicative of the effectiveness of the DD strategy space exploration. A simulation demonstrates that in 7 GA iterations (Fig. \ref{fig:unif_start}), more of the DD strategy space is explored starting from the uniform population than a random starting population in expectation at any mutation probability and higher mutation probabilities lead to smaller differences between the behaviors of the two in terms of number of unique strategies explored as expected. This benefit is particularly important when a real optimization landscape with DD utilities is imposed, as exploring a larger fraction of the parameter space at any iteration number leads to lower probabilities of converging to local extrema that are suboptimal compared to a globally optimal DD strategy.

In each iteration of the genetic algorithm, $K$ pairs of parent DD strategies are selected with replacement from the $K$ individuals in the previous generation for reproduction. The probability that any individual is selected for reproduction is proportional to 
\begin{equation}
    \log{10u + 1}
\end{equation}
where $u$ is the utility function of the observed count distribution associated with a DD strategy normalized to $[0,1]$. Following the construction in Ref.~\cite{Qui13}, this has the goal of converging towards the highest utility DD strategies, due to both the survival of high utility DD strategies across GA iterations and the increased probability of selecting high utility DD strategies for reproduction. After reproduction to create $2K$ offspring from the $K$ pairs of parent strategies, the utility function is re-computed for the $K$ elements of the previous population and the $2K$ offspring, and $K/4$ highest utility parents and $3K/4$ highest utility offspring are chosen for the new population to conserve the overall population size $K$. 

The reproduction and mutation algorithms are also developed from generalizing the protocols described in Ref.~\cite{Qui13} to the multi-qubit regime. For any pair of length $L$ single-qubit pulse sequences, the reproduction protocol is defined 
\begin{align}
\begin{split}
    & c_{11}c_{12}\dots c_{1L}  \\
    & c_{21}c_{22}\dots c_{2L} 
\end{split}
\end{align}
by selecting a random site $l \in \{1, 2, \dots, L\}$. Then, the offspring pulse sequences can be written as
\begin{align}
\begin{split}
    c'_{11}c'_{12}\dots c'_{1L} &\equiv c_{11}c_{12}\dots c*_{1l}c_{2(l+1)}c_{2(l+2)}\dots c_{2L} \\
    c'_{21}c'_{22}\dots c'_{2L} &\equiv c_{21}c_{22}\dots c*_{2l}c_{1(l+1)}c_{1(l+2)}\dots c_{1L},
\end{split}
\end{align}
where $c*_{1l}$ and $c*_{2l}$ are selected to be the unique elements of $G$ that make the resulting sequences multiply to the identity. Furthermore, we implement the single-pulse where for each offspring $c'$, where two pulse locations $1 \leq l, l' \leq L$ are selected. Then, $c'_l$ is mutated to some element of $G$ selected uniformly a t random with a fixed probability, and $c'_{l'}$ is then constrained to again make the resulting sequence multiply to the identity. This protocol is extended to DD strategies on $C$ qubit colors by performing single-qubit reproduction and mutation on each color, followed by a randomized assignment of the two single-color offspring to the two offspring strategies.

For each new strategy generated by reproduction, the single pulse mutation occurs with some probability that is fixed per iteration but can vary throughout iterations of the genetic algorithm. This dynamic variation of the mutation probability is designed to accelerate the convergence of the genetic algorithm to the desired result. We find good convergence in the aforementioned results by initially setting the mutation probability to around $0.7$ and dynamically increasing or decreasing the mutation probability by $0.1$ if the system appears to be far away or close to equilibrium respectively. Due to the complicated DD optimization landscape, the state of being in or out of equilibrium is difficult to quantify via iteration snapshots. We find that using proxy quantities such as the standard deviation or range of utility functions exhibited by all individuals in the population at any individual GA iteration gives reasonably good convergence results. However, such a condition is significantly dependent on the problem structure and utility function selection and there has not been sufficient rigorous testing for attribution of varying mutation rates to varying convergence.

\section{Cliffordized circuits and Grover's algorithm \label{ss:clifford}}

\begin{figure*}
    \centering
    \includegraphics[height=5cm]{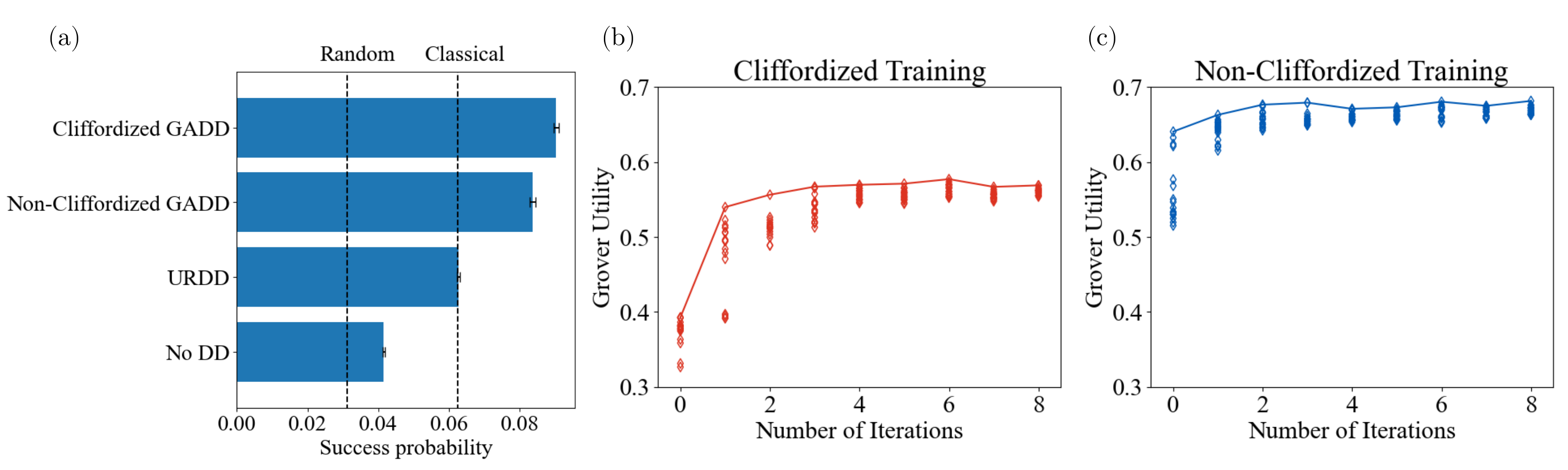}

    \caption{(a) 5-qubit Grover's algorithm's success probabilities averaged over all $2^5$  oracles and the utility vs. iteration for GADD training on (b) Cliffordized and (c) non-Cliffordized, or the original, Grover circuits on $\texttt{ibmq\_mumbai}$. Here, the GADD strategy was trained with the utility function relating to the difference from the ideal distribution as classical simulated. In both the Cliffordized and non-Cliffordized training cases, we see the convergence of the GA within $\sim 5$ iterations. Iterations $9-20$ provide more iterations where the population has simply converged and is thus omitted. We note that for the non-Cliffordized circuit, our initial strategy population happened to contain high-utility individuals. The average success probability in (a) of GADD strategies from Cliffordized and non-Cliffordized training is higher than the URDD strategy and no DD strategy. The random threshold is established from a simple random uniform guess overall $2^5$ bitstrings, while the best classical strategy allows for a single classical query before guessing.}
    \label{fig:grover_conv}
    \label{fig:grover training}
\end{figure*}
We implement Cliffordization to preserve DD motifs present in a target circuit while reducing it to a training circuit that is efficiently classically simulable. Generically, for quantum gate $U$ imparting phase $0 \leq \phi \leq \frac{\pi}{2}$, $\phi$ is set to $\frac{\pi}{2}$ with probability $\frac{\phi}{\pi/2}$ and $0$ with the remaining probability. Analogously, this conversion is performed for $\phi \in (\frac{\pi}{2}, \pi), (\pi, \frac{3\pi}{2}), (\frac{3\pi}{2}, 2\pi)$ with their respective bounds. If $U$ also imparts a bit flip, the bit flip is converted into a phase rotation by the Hadamard gate, the previous protocol is applied to the phase rotation, and the resulting Clifford phase rotation is converted back into a bit flip with another Hadamard gate. For example, the $T^{\dag}$ gate imparting a $\frac{-\pi}{4}$ phase on a quantum state in the computational basis is Cliffordized by replacing it with a gate with $\phi = \frac{3\pi}{2}$ or the identity with $\phi = 0$ with equal probability. 

The Grover's search algorithm provides an ideal platform to evaluate the concept of Cliffordized training. Grover's search algorithm seeks to find a hidden length $N$ bitstring $b$ in an unsorted list with $2^N$ elements. The oracle in Grover's algorithm returns $f_b(x) = \delta_{x,b}$ when queried with the input bitstring $x$. Classically, searching such an unsorted list requires $O(N)$ queries. By encoding the problem in an $N$ qubit circuit, a quadratic improvement in the number of queries is provably achieved relative to the best possible classical algorithm \cite{Gro97}. However, even for relatively small $N$, the circuit depths associated with even the most efficient implementations of Grover's algorithm do not exhibit better-than-classical results without DD due to errors \cite{Pok22}.

Grover's search algorithm provides an ideal platform to evaluate the concept of Cliffordized training for two reasons. First, there are $2^N$ Grover search problems on $N$ qubits that can be easily accessed by changing specific parameters in non-Clifford gates present in the Grover oracle, but without changing the general circuit structure. Second, the depth of the Grover circuit imposes that although the $N = 5$ case poses challenges in terms of error on the quantum device, it is sufficiently small for individual non-Clifford Grover circuits to be tractably simulated classically. As a result, the effects of training with a Cliffordized circuit can be effectively compared with the effects of training with the circuit before Cliffordization.

We begin by considering two different training circuits for GADD. First, we consider the Cliffordized version of the Grover circuit encoding target bitstring $11111$, as well as its original non-Cliffordized form as a control to evaluate the effect of Cliffordization. For each of these circuits, we run the GADD algorithm for DD strategies with the same hyperparameters as in the BV problem, except with utility function $u(t)$ which considers total absolute difference between the simulated and experimentally estimated distribution of output counts in the computational basis with $N = 5$ qubits as described in \cref{eqn:1norm_utility} over $4000$ shots. We note that this utility function has no knowledge of what the Grover target string is in the Cliffordized case. The resulting learned DD sequence is then padded to all $32$ distinct circuits for $N = 5$ Grover search problems and the average search success probability over the $32$ circuits is calculated. These circuits are run on the 27-qubit \verb|ibmq_mumbai| backend, which is a superconducting qubit device with the IBM Falcon architecture.

In both the Clifford and non-Clifford training cases, we observe convergence in the population generated by GADD after $\sim 4$ GA iterations (\cref{fig:grover_conv}). After implementing these learned high-utility DD strategies on all $N = 5$ Grover circuits, we demonstrate that significantly better-than-classical success probabilities are achieved under the implementation of GADD under both training cases as depicted in the right panel of Fig.~\ref{fig:grover training}. In contrast, this is not achieved in the case where URDD, the strategy that generically gave the next highest performance in Sec.~\ref{sec:bv}, is implemented. Notably, the Grover success probability is comparable in the Cliffordized and non-Cliffordized training cases, which provides empirical evidence for the effectiveness of GADD training on a result-agnostic utility function. As the Cliffordized circuit provides a more general representation of the $N = 5$ qubit Grover problem overall Grover oracles, while the non-Clifford circuit preserves the non-generalizable information of $11111$ being the target string, the comparable resulting Grover success probabilities imply that encoding such output-dependent information in the GADD utility function is unimportant. Thus, the Cliffordization method for generating training circuits is quite promising for learning successful DD strategies in the general setting where the resulting quantum state from the problem at hand is \textit{a priori} unknown.

\section{Hardware and software specifications}
\label{app:device}

\begin{table*}[t]
    \centering
    \begin{tabular}{|l|c|c|c|c|c|c|c|c|c|c|c|c|}
    \hline
        ~ & \multicolumn{2}{c}{Peekskill (BV)}  & & \multicolumn{2}{c}{Mumbai (Grover v1)} & & \multicolumn{2}{c}{Mumbai (Grover v2)}  & & \multicolumn{2}{c}{Kyiv (MRB)} &   \\ \hline
        ~ & Min & Mean & Max & Min & Mean & Max & Min & Mean & Max & Min & Mean & Max \\ \hline
        $T_1$ ($\mu$s)  & 139.66 & 321.63 & 516.35 & 57.32  & 113.65  & 174.98 & 21.17 & 141.42 & 343.44 & 37.4 & 277.7 & 465.4 \\ \hline
        $T_2$ ($\mu$s) & 23.66  & 299.26 & 536.11 & 27.06 &  167.27 & 343.3 & 27.06  & 167.27 & 343.3 & 4.0 & 120.3 & 506.0 \\ \hline
        1QG Error (\%) & 0.01 & 0.02 & 0.05 & 0.01 & 0.03 & 0.19 & 0.01 & 0.03 & 0.1 & 0.01 & 0.03 & 0.49 \\ \hline
        2QG Error (\%) & 0.39  & 0.8 & 1.37 & 0.49 & 0.91 &  2.1 & 0.49  & 0.91 & 1.76 & 0.37 & 1.1 & 100. \\ \hline
        1QG Duration ($\mu$s) & 0.04 & 0.04 & 0.04 & 0.04 & 0.04 & 0.04 & 0.04 & 0.04 & 0.04 & 0.05 & 0.05 & 0.05 \\ \hline
        2QG Duration ($\mu$s) & 0.43 & 0.6 & 0.62 & 0.25 &  0.45 &  0.74 & 0.25 & 0.45 & 0.74 & 0.56 & 0.56 & 0.56 \\ \hline
        RO Error (\%) & 0.37 & 1.97 & 9.59 & 1.0 &  2.84 &  8.02 & 1. & 2.84 & 8.02 & 0.14 & 0.88 & 28.82 \\ \hline
        RO Duration ($\mu$s) & 0.86 & 0.86 & 0.86 & 3.58 & 3.58 & 3.58 & 3.58 & 3.58 & 3.58 & 1.24 & 1.24 & 1.24 \\ \hline
    \end{tabular}
\caption{Device specifications for \texttt{ibm\_peekskill}, \texttt{ibmq\_mumbai} and \texttt{ibm\_kyiv} at the time of their respective experiment executions are shown. Grover v1 and v2 refer to the Cliffordized and non-Cliffordized runs respectively. 1QG, 2QG, and RO denote 1-qubit gate, two-qubit gate, and readout, respectively.}
\label{tab:config}
\end{table*}

\begin{table*}[t]
    \centering
    \begin{tabular}{|l|c|c|c|c|c|c|c|c|c|}
    \hline
    ~ & \multicolumn{3}{c|}{Sherbrooke (GHZ, 2024-06-13)}
      & \multicolumn{3}{c|}{Sherbrooke (GHZ, 2023-06-19)}
      & \multicolumn{3}{c|}{Kyiv (GHZ, 2024-06-13)} \\ \hline
    ~ & Min & Mean & Max & Min & Mean & Max & Min & Mean & Max \\ \hline
$T_1$ ($\mu$s) & 23.19 & 265.79 & 475.14 & 50.15 & 257.87 & 484.34 & 33.34 & 277.95 & 465.45 \\ \hline
$T_2$ ($\mu$s) & 15.64 & 172.51 & 394.96 & 16.93 & 174.29 & 516.61 & 15.52 & 153.08 & 542.36 \\ \hline
1QG Error (\%) &  0.01 &   0.04 &   0.62 &  0.01 &   0.04 &   0.62 &  0.01 &   0.04 &   0.67 \\ \hline
2QG Error (\%) &  0.33 &   5.13 & 100.00 &  0.31 &   5.04 & 100.00 &  0.34 &    1.3 &    6.3 \\ \hline
1QG Duration ($\mu$s) &  0.06 &   0.06 &   0.06 &  0.06 &   0.06 &   0.06 &  0.05 &   0.05 &   0.05 \\ \hline
2QG Duration ($\mu$s) &  0.34 &   0.54 &   0.88 &  0.34 &   0.54 &   0.88 &  0.56 &   0.56 &   0.56 \\ \hline
RO Error (\%) &  0.27 &   2.79 &  34.77 &  0.35 &   2.68 &  38.16 &  0.07 &   1.64 &  19.97 \\ \hline
RO Duration ($\mu$s) &  1.24 &   1.24 &   1.24 &  1.24 &   1.24 &   1.24 &  1.24 &   1.24 &   1.24 \\ \hline
    \end{tabular}
\caption{Device specifications for \texttt{ibm\_sherbrooke} and \texttt{ibm\_kyiv} at the time of experiment execution for GHZ state preparation experiments are shown. 1QG, 2QG, and RO denote 1-qubit gate, two-qubit gate, and readout, respectively.}
\label{tab:config_ghz}
\end{table*}
All experimental results in this work were obtained through Qiskit \cite{Qiskit}. The code for MRB circuits and analysis was written in Qiskit Experiments \cite{Qiskit_Experiments}. To quickly iterate in the genetic algorithm training process, Qiskit Runtime \cite{qiskit_runtime} was used to interleave quantum training circuits and the classical learning algorithm to evaluate the strategies.

BV, Grover, and MRB experiments were run on the following IBM superconducting quantum processors respectively: \texttt{ibm\_peekskill} and \texttt{ibmq\_mumbai} are 27-qubit Falcon r5.10 and r8 processors, respectively; \texttt{ibm\_kyiv} is a 127-qubit Eagle r3 processor. The machine specifications for all three devices for the data reported in this work are shown in~\cref{tab:config}. GHZ experiments were run on \texttt{ibm\_kyiv} and \texttt{ibm\_sherbrooke}, another 127-qubit Eagle r3 processor, with device specifications shown in ~\cref{tab:config_ghz}. Note that in certain cases the maximum two-qubit errors are at 100\% due to certain qubits being out-of-commission; in such cases, we transpiled our circuit to avoid those qubits.

\end{document}